%% file: complete_draft_v8.tex
\documentclass[preprint]{elsarticle}

\pdfoutput=1                   % for ensuring arxiv uses pdflatex

\usepackage{hyperref}    % for making the table of contents clickable
\hypersetup{linktoc=all}

\usepackage{titlesec}     % for changing the format of my titles
\usepackage[titletoc]{appendix}    % for adding nice appendix title to the table of contents

\titleformat{\paragraph}
{\normalfont\normalsize\itshape}{\theparagraph}{1em}{}
\titlespacing*{\paragraph}
{0pt}{3.25ex plus 1ex minus .2ex}{1.5ex plus .2ex} % To get a usuable paragraph command

\usepackage{graphicx}    % for handling figures in the document

\usepackage{tabularx}    % for handling tables with variable width columns and text wrapping
\usepackage{tabulary}     % for handling tables with variable width columns and text wrapping

\usepackage{array}
\newcolumntype{K}[1]{>{\raggedright\let\newline\\\arraybackslash\hspace{0pt}}m{#1}}    %for having a single variable width column

\usepackage{booktabs}    % for making tables pretty
\usepackage[justification=centering]{caption}

\usepackage[font=normalsize, skip=5pt]{caption}    % for customizing captions

\usepackage{amsmath}   % for math environments
\usepackage{amssymb}   % for special math symbols

\usepackage{color, soul}    % for highlighting

\biboptions{round, semicolon, authoryear}

\usepackage{url}    %for handling of urls in bibtex file
\usepackage{pdflscape}    % for landscape orientation of pages

\usepackage{verbatim}    % for multi-line comments

\usepackage[margin=1in]{geometry}    % to get 1-inch margins

\usepackage{todonotes}    % For adding notes to the margin

\begin{document}
\begin{frontmatter}

\title{Asymmetric, Closed-Form, Finite-Parameter Models of Multinomial Choice}
\date{November 22, 2016}

\author[tbrathwaite]{Timothy Brathwaite\corref{cor1}}
\ead{timothyb0912@berkeley.edu}

\author[jwalker]{Joan Walker}
\ead{joanwalker@berkeley.edu}

\cortext[cor1]{Corresponding Author}

\address[tbrathwaite]{Department of Civil and Environmental Engineering, University of California at Berkeley\\ 116 McLaughlin Hall, University of California, Berkeley, CA, 94720-1720}
\address[jwalker]{Department of Civil and Environmental Engineering, University of California at Berkeley\\ 111 McLaughlin Hall, University of California, Berkeley, CA, 94720-1720}

\begin{abstract}
Class imbalance, where there are great differences between the number of observations associated with particular discrete outcomes, is common within transportation and other fields. In the statistics literature, one explanation for class imbalance that has been hypothesized is an asymmetric (rather than the typically symmetric) choice probability function. Unfortunately, few relatively simple models exist for testing this hypothesis in transportation settings---settings that are inherently multinomial. Our paper fills this gap.

In particular, we address the following questions: ``how can one construct asymmetric, closed-form, finite-parameter models of multinomial choice'' and ``how do such models compare against commonly used symmetric models?'' Methodologically, we introduce (1) a new class of closed-form, finite-parameter, multinomial choice models, (2) a procedure for using these models to extend existing binary choice models to the multinomial setting, and (3) a procedure for creating new binary choice models (both symmetric and asymmetric). Together, our contributions allow us to create new asymmetric, closed-form, finite-parameter multinomial choice models. We demonstrate our methods by developing four new asymmetric, multinomial choice models. Empirically, most of our models strongly dominate the multinomial logit (MNL) model in terms of in-sample and out-of-sample log-likelihoods. Moreover, analyzing two policy applications, we find practical differences between the MNL and our new asymmetric models. Our results suggest that while asymmetric models may not always outperform symmetric ones, asymmetric choice models are worth testing because they might have better statistical performance and entail substantively different policy and financial implications when compared with traditional symmetric models, such as the MNL.
\end{abstract}

\end{frontmatter}

\section{Introduction}
\label{sec:intro}
Discrete choice modeling is widely used in transportation. It is used in every area of travel demand analysis, such as residential choice, work location choice, destination choice, time-of-travel choice, mode choice, and route choice. Moreover, discrete choice modeling is also used outside of transportation in  fields such as marketing, economics, finance, operations research, statistics, and medicine. Across these many disciplines, the most commonly used models have fairly simple functional forms, such as the multinomial logit (MNL) and binary logit models. The use of simple models is, in part, due to the greater computational burdens required to estimate and forecast with very general discrete choice models. Clearly then, it is important to create simple models that are nonetheless able to avoid unwanted properties of classic models such as the MNL model. In this paper, we introduce models that have the same basic form as the MNL model but, for the price of a finite number of new parameters that are to be estimated from the data, provide potentially much better fits to one's data and avoid a ``symmetry property'' that we argue is often undesirable. The next paragraph will review the MNL model because it is the starting point for the class of models that we introduce. Then, we will describe the symmetry property, show it is present in common discrete choice models, and make the case that such a property is not always desirable.

While the MNL and binary logit models are often used because of their ease of estimation, their closed-form probability equations (shown in Equation \ref{eq:mnl_formula})\footnote{Note that the variables are fully listed in order to make clear the notation used in the paper.}, and their ease of interpretation, their use requires analysts to accept a set of properties that may be overly restrictive or inaccurate in the specific contexts being modeled. Specifically, one well known property is known as Independence from Irrelevant Alternatives (I.I.A.). The I.I.A. property is seen as problematic when one considers substitution patterns between alternatives that are closely related, and there have been numerous models that aim to avoid I.I.A. (e.g. nested logit, cross-nested logit, etc.).

\begin{equation}
\label{eq:mnl_formula}
\begin{aligned}
P \left( y_{ij} = 1 | V_{i1}, V_{i2}, ..., V_{ik} \ \forall \left\lbrace j, k \right\rbrace \in C_i \right) &= \frac{\exp \left( V_{ij} \right)}{\sum _{\ell \in C_i} \exp \left( V_{i \ell} \right)} \\
\textrm{where } y_{ij} &= \textrm{a binary (0 or 1) indicator of whether individual $i$} \\
&\quad \  \textrm{is associated with outcome $j$.} \\
C_i &= \textrm{the choice set for individual $i$} \\
V_{ij} &= x_{ij} \beta = \textrm{the index for alternative $j$ for individual $i$} \\
\beta &= \textrm{a column vector of unknown population parameters} \\
x_{ij} &= h \left( z_j, \zeta _i \right) \textrm{, a row vector.} \\
h \left(  \right) &= \textrm{a function that returns a row vector} \\
z _j &= \textrm{attributes of alternative $j$ for individual $i$} \\
\zeta _i &= \textrm{characteristics of individual $i$}
\end{aligned}
\end{equation}

\begin{figure}
\centering
\includegraphics[width=\textwidth]{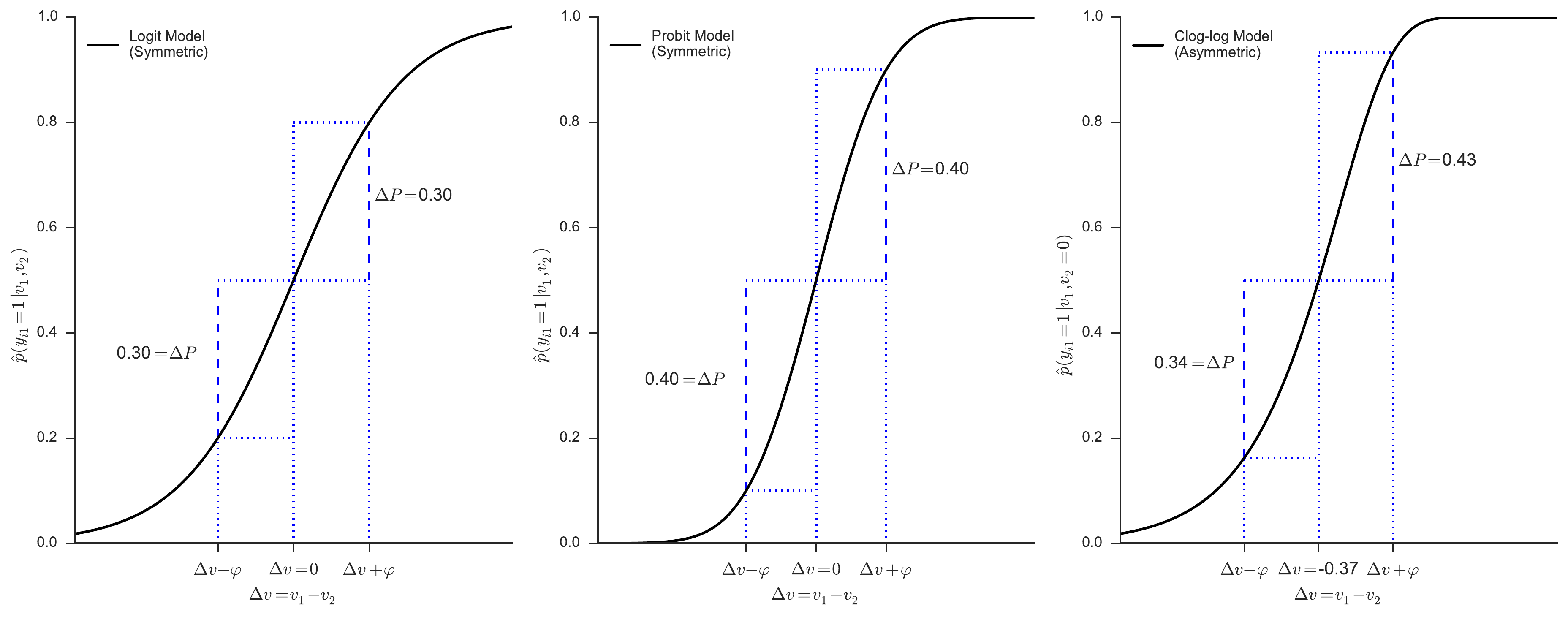}
\caption{Symmetric and Asymmetric Binary Choice Probability Functions}
\label{fig:symm_logit_probit}
\end{figure}

In addition to the I.I.A. property, the MNL model's probability function also implies a ``symmetry property.'' Specifically, from a point where an individual has a 50\% probability of choosing an alternative $j$, this probability will increase and decrease at equal rates with respect to equal-magnitude increases and decreases in alternative $j$'s index, $V_{ij}$. Probability functions with this quality are henceforth referred to as symmetric, and probability functions without this property are henceforth referred to as asymmetric. See Figure \ref{fig:symm_logit_probit} for a visual depiction of symmetric and asymmetric probability functions. The binary, complementary log-log model (henceforth clog-log model) is described in Section \ref{sec:deriving_multinomial_cloglog} and used in Figure \ref{fig:symm_logit_probit} as an example of an asymmetric probability function. In contrast, the binary logit and binary probit models are used as examples of symmetric probability functions. Note that the logit model is not the only model with the symmetry property. The other commonly used discrete choice model, the simple probit model\footnote{The `simple' probit model assumes that the error terms of the utility of each alternative are independent and identically distributed.}, is also symmetric. The point being made here is that while it is seldom spoken of, a basic property of standard discrete choice models is that one's probability of choosing a given alternative is symmetric about 50\%, with respect to the index, $V_{ij}$, of that alternative.

Although models exhibiting the symmetry property are pervasive in discrete choice modeling, there are situations where such a property may seem overly restrictive. Class-imbalanced choice contexts, where the numbers of observations choosing each alternative are unequal, are one such set of situations. Note that in transportation, class-imbalanced choice contexts are ubiquitous. For example, in the United States (US), there are almost always many more automobile drivers than bicyclists when modeling commute mode choices. Furthermore, while the initial motivation and focus of our empirical applications is on travel mode choice, we emphasize that class imbalance is prevalent in many settings where discrete choice models are employed. For instance, class imbalance is observed in: biomedical studies of the dose-response effects of drugs \citep{pregibon_goodness_1980}; destination choice studies in the form of 'superstar' destinations \citep{chorus_paving_2016}; loan default studies \citep{calabrese_modelling_2013}; and studies of shoppers' brand choice \citep{briesch_semiparametric_2002}.

In class-imbalanced situations, it might be natural to hypothesize that the probability of choosing the under-represented alternative decreases more rapidly from 50\% than it increases, even for equal-magnitude decreases and increases in the alternative's index, $V_{ij}$. Of course, this hypothesis is not the only plausible explanation for the observed class imbalance\footnote{As noted by one referee, there are numerous methods in use in statistics and machine learning for ameliorating the effects of class imbalance on one's chosen performance metric. For example, there exist many types of over- and under-sampling techniques. These techniques deal with the effect of class imbalance on \textit{prediction}. The focus of this paper is different. We use asymmetric probability models to accommodate alternative \textit{explanations} of why class imbalance is observed.}. The point, however, is that symmetric probability models prohibit one from investigating hypotheses about the magnitude of changes in the probability of choosing an alternative, from a probability of 50\%, with respect to equal-magnitude increases and decreases in that alternative's index. This is because symmetric probability models assume \textit{a-priori} that the changes in probability are equal. 

In light of this undesired symmetry property of common discrete choice models such as the standard MNL and simple probit model, this paper's contributions to the transportation and discrete choice literature are that it:
\begin{enumerate}
\item introduces a general class of closed-form, finite-parameter models for multinomial choice situations that do not necessarily imply symmetric choice probability functions (as well as four new models within that class),

\item introduces and demonstrates a methodology for
    \begin{enumerate}
        \item extending existing, binary choice models to the multinomial setting and
        \item creating new binary choice models (both asymmetric and symmetric),
	\end{enumerate}

\item demonstrates that asymmetric probability models can substantially improve upon the fit of standard discrete choice models such as the MNL model, and
 
\item shows that, compared to symmetric models such as the MNL model, asymmetric choice probability functions can lead to substantive differences (both quantitatively and qualitatively) in one's resulting statistical inference and policy-analyses.
\end{enumerate}

For clarity, we reiterate the main purpose of this paper. First, we note that class imbalance is a common occurrence in discrete choice analyses. Secondly, through extensive reference to the existing statistical literature, we highlight the fact that asymmetric probability functions have been given as \textit{\textbf{one possible}} explanation for why there might be low relative numbers of individuals choosing a particular alternative. Another possible explanation is a data-generating process with a symmetric probability function and low, average systematic utilities in one's population for the under-represented alternatives, relative to the over-represented alternatives. In general, we do not think that class imbalance \textit{\textbf{necessarily}} implies an asymmetric probability function. Moreover, we do not think that asymmetric probability functions will \textit{\textbf{necessarily}} perform better than symmetric ones when modeling class imbalanced data. Investigation of such claims are beyond the scope of this paper. Through reference to the existing statistical literature, and through our empirical applications, we instead demonstrate that asymmetric probability functions \textit{\textbf{can possibly}} provide better explanations of the observed choices in one's class imbalanced dataset, and that due to this possibility, one should investigate the use of asymmetric probability functions in one's analyses. To facilitate the use of asymmetric probability functions in discrete choice analyses, we create methods to construct new, binary probability functions (both symmetric and asymmetric), and we create methods to extend binary probability functions to the multinomial settings that are common in many fields that use discrete choice models.

The rest of the paper is organized as follows. Section \ref{sec:lit_review} will review related literature and current approaches to producing discrete choice models that are not necessarily symmetric. Section \ref{sec:logit_type_models} will detail our proposed class of choice models, relate it to the existing literature, and show how one might create such models. Section \ref{sec:model_estimation} will describe the estimation procedures for our proposed models, and Section \ref{sec:empirical_applications} will detail our empirical examples and case studies, comparing our proposed models to existing ones such as the MNL model. Section \ref{sec:extensions} will discuss extensions of our work and Section \ref{sec:conclusion} will conclude.

\section{Literature Review}
\label{sec:lit_review}
One can partition the asymmetric discrete choice models that have been proposed in the literature based on whether they:
\begin{itemize}
\item are binary or multinomial choice models,
\item are closed- or open-form\footnote{The probability equation of open-form models contain analytically intractable integrals or infinite sums.} models,
\item have a null, finite, or infinite\footnote{Models with an infinite number of parameters are known as non-parametric or semi-parametric models.} set of shape parameters---i.e. parameters that control the shape of the resulting choice probability function.
\end{itemize}
To review the literature that this paper builds upon, we will iterate through each of these descriptors in the coming paragraphs---describing the work that has been done so far, how that work relates to or has been used in transportation, and issues with the existing literature that our paper addresses.

First, virtually all research that explicitly focuses on the development of asymmetric choice models has been carried out in the binary setting. Since at least 1976, statisticians and computer scientists have been introducing closed-form, asymmetric generalizations of the standard binary logit model through the use of one or two shape parameters \citep{prentice_generalization_1976, pregibon_goodness_1980, aranda-ordaz_two_1981, guerrero_use_1982, stukel_generalized_1988, morgan_extended_1988, czado_link_1992, czado_parametric_1994, nagler_scobit:_1994, chen_new_1999, vijverberg_betit:_2000, masnadi-shirazi_variable_2010, vijverberg_pregibit:_2012, nakayama_unified_2015, komori_asymmetric_2015}. These shape parameters allow one to adapt the shape of the resulting choice probability function to fit the data at hand. Beyond generalizations of the logit model, a number of binary, asymmetric models that do not nest the logit model have also been proposed in the statistics literature. For example, the clog-log model has been around since at least the 1920s \citep{fisher_mathematical_1922, yates_use_1955, mccullagh_generalized_1989}, and the GEV regression model (not to be confused with McFadden's GEV distribution) is a generalization of the clog-log model with one shape-parameter \citep{wang_generalized_2010, calabrese_modelling_2013}. Still other asymmetric models of binary choice have been introduced based on skewed normal distributions \citep{bazan_framework_2010}, skewed student's t-distributions \citep{kim_binary_2002, kim_flexible_2008}, and symmetric power distributions \citep{jiang_new_2013}. Broadly, the binary choice models with one or more shape parameters have been referred to as ``parametric link functions'' in the statistics literature \citep{mccullagh_generalized_1989}, and many examples of asymmetric choice models can be found by searching for scholarly articles that use such phrases.

Two problems exist with the asymmetric choice models just discussed. First, while the litany of binary, asymmetric choice models that has been developed may be quite useful, they must be extended to the multinomial setting for use in transportation contexts---contexts where the choice situations are often inherently multinomial. Secondly, the proliferation of binary, asymmetric choice models suggests that no single asymmetric model fits the needs of all researchers. However, no guidance on how to create such asymmetric models has been offered in the literature. The various models cited in the last paragraph were almost all introduced without any explanation of where the functional form for the model came from. Sections \ref{sec:logit_type_models} resolves these issues by detailing a method for extending binary, asymmetric models to the multinomial setting and by introducing a methodology for creating binary, asymmetric choice models.

In addition to all of the binary, closed-form, asymmetric choice models described above, many binary, asymmetric choice models with open-form choice probability functions have also been proposed. These open-form models are typically one of two major varieties. One type of binary, open-form, asymmetric model uses an asymmetric probability density function for the difference in the error terms of the utilities of the two alternatives. Examples of this type include the aforementioned models that were based on the skewed normal distributions \citep{bazan_framework_2010} and skewed student's t-distributions \citep{kim_binary_2002, kim_flexible_2008}. The second type of binary, open-form, asymmetric model is based on a mixed logit or mixed probit approach, whereby a random variable with an asymmetric probability density function is added to the index, $V_{ij}$, of the alternative of interest. In this second type of model, the random variable with an asymmetric probability density function is multiplied by an unknown coefficient whose value is to be estimated. If the estimated coefficient's value is zero, then the model reduces to the symmetric probability model (e.g. logit or probit) being used as the kernel of the asymmetric model. Examples of this type of model include the bayesian asymmetric logit and bayesian asymmetric probit models \citep{chen_new_1999}.

The first type of binary, open-form, asymmetric model described above might be easily extended to the multinomial setting, provided that there exist multivariate versions of the asymmetric probability density functions that are used in the binary case, or provided that such multivariate distributions can be created. This remains an open question. On the other hand, the second type of binary, open-form, asymmetric model can be easily extended to the multinomial setting by simply adding random variables with asymmetric probability density functions to each of the utility functions for the alternatives in one's model. However, regardless of whether such models can be extended to handle multinomial choice situations, such open-form models will still entail computational burdens in estimation, storage, and forecasting, relative to their closed-form counterparts. In this paper, we focus on developing closed-form, asymmetric choice models because they are less computationally burdensome and more closely parallel the discrete choice models that have been used in all industries (namely closed-form models such as the MNL model).

Regarding the number of shape parameters in one's model, all of the asymmetric choice models that have been mentioned so far have had one or two shape parameters, with the exception of the binary clog-log model. In contrast to this, many multinomial, asymmetric choice models have been inadvertently\footnote{We say inadvertently because in none of the cases cited was the purpose of creating the model to avoid the symmetry property discussed in Section \ref{sec:intro}.} created by transportation researchers, and most of them have no shape parameters. Unlike the binary, closed-form, asymmetric models discussed above---where the functional form of $P \left( y_{i1} = 1 \mid V_{i1}, V_{i2} \right)$ is assumed outright---the multinomial, asymmetric models created by transportation researchers come from assuming various distributions for the error terms in the utility equations for each alternative. In particular, multinomial, asymmetric choice models have been derived by transportation researchers by assuming Weibull \citep{castillo_closed_2008, fosgerau_discrete_2009}, Rayleigh \citep{li_multinomial_2011}, Type II Generalized Logistic \citep{li_multinomial_2011}, Pareto \citep{li_multinomial_2011, mattsson_extreme_2014}, Exponential \citep{li_multinomial_2011}, and Fr{\'e}chet \citep{mattsson_extreme_2014} distributions for the utilities of one's alternatives. In each of these cases, the resulting multinomial choice model is asymmetric. A more recent paper \citep{nakayama_unified_2015} uses a ``$q$-GEV'' distribution for the utility of each alternative, and derives a multinomial, asymmetric choice model with one shape parameter, $q$. As a brief aside, Li (\citeyear{li_multinomial_2011}) and Nakayama and Chikaraishi (\citeyear{nakayama_unified_2015}) note that all of the models described in this paragraph have probability equations with the same functional form as the MNL model, except that $V_{ij}$ is replaced with $S_{ij}$. Here, $S_{ij} = S \left(V_{ij}, \gamma _j \right)$ or $S_{ij} = S \left(V_{ij} \right)$ depending on whether the model has shape parameters ($\gamma _j$), and $S \left( \cdot \right)$ is a monotonically increasing function of $V_{ij}$. Note also, that the one multinomial, closed-form, asymmetric model that has been introduced in the statistics literature \citep{das_generalized_2014} also has this form, except that $\gamma _j = \left[ \gamma_{j1}, \gamma_{j2} \right]^T$, i.e. there are two shape parameters per alternative. This functional form will be mentioned again in Section \ref{sec:logit_type_models} as it is very similar to the one that we propose in this paper.

While all of the asymmetric models introduced by transportation researchers share the virtue of being able to handle multinomial choice situations, they all share a key drawback: they are only valid for certain values of the index, $V_{ij}$. To be concrete, the weibit model of Castillo et al. (\citeyear{castillo_closed_2008}) and Fosgerau and Bierlaire (\citeyear{fosgerau_discrete_2009}) is only defined for values of $V_{ij}$ that are negative. The same is true for utility maximizing models based on the Rayleigh, Type II Generalized Logistic, or Exponential distributions \citep{li_multinomial_2011}. If the model is based on the Pareto distribution, then $V_{ij}$ must be less than negative one \citep{li_multinomial_2011, mattsson_extreme_2014}, and if the model is based on the $q$-GEV distribution, then $V_{ij}$ must be greater than or equal to $\frac{-1}{q - 1}$ for whatever value of $q$ is specified or estimated from one's data \citep{nakayama_unified_2015}. In choice situations where the index, $V_{ij}$, should be comprised of both variables that increase an alternative's probability of being chosen and variables that decrease an alternative's probability of being chosen, it can be hard or impossible to meet such constraints on the index's value or sign. Because of this, the models mentioned in the last paragraph are only applicable in a restrictive set of circumstances. In Section \ref{sec:logit_type_models}, we will introduce a class of multinomial, closed-form, asymmetric choice models that is (in general) free from the sign and magnitude restrictions on  $V_{ij}$ that have limited the usefulness of asymmetric choice models in transportation so far. Our class of models will be shown to include the previously derived models as special cases.

Lastly, transportation researchers in the multinomial setting \citep{li_multinomial_2011}, and econometricians in the binary setting \citep{horowitz_semiparametric_1993}, have specified closed-form, asymmetric choice models that have an infinite number of shape parameters. That is to say, closed-form, asymmetric models have been specified where the function $S \left( V_{ij} \right)$, as defined above, has been estimated non-parametrically. These models are known in the econometrics literature as single-index models \citep{hardle_semiparametric_1997, horowitz_semiparametric_2010}. As shown by Li (\citeyear{li_multinomial_2011}), single-index models can take on symmetric or asymmetric forms. While these models are quite general, and they avoid the problems that come from mis-specifying one's choice probability function \citep{czado_effect_1992, koenker_parametric_2009}, they can be difficult to estimate and require rather large sample sizes to estimate with decent precision. For these reasons, we develop a class of models in Section \ref{sec:logit_type_models} that depends on a finite number of shape parameters, making the class more flexible than the fixed shape models that are classically used in transportation such as MNL models but less computationally burdensome than the single-index models described above.

\subsection{Summary}
\label{sec:lit_review_summary}
Overall, across a variety of academic disciplines, many asymmetric choice models have been created thus far. However, this development has been fragmented and leaves much room for improvement. In particular, most of the existing asymmetric models are binary models, but to be most useful in transportation, these binary models need to be extended to the multinomial setting. Moreover, we need systematic methods for creating new asymmetric models when the existing ones do not meet our research needs. In the previous literature, there has been much work on creating asymmetric, open-form, binary choice models. In this paper, we do not pursue the development of such models because of their greater computational complexity in estimation, storage, and forecasting in comparison to their closed-form counterparts. For the same reason, we do not consider closed-form, multinomial, asymmetric models with an infinite number of shape parameters. Instead, we build on the work of transportation researchers since they have created numerous multinomial, asymmetric models that have zero or a finite number of shape parameters. A major limitation of the asymmetric models in transportation is that they all restrict the values that the index, $V_{ij}$, can take. In the next section, we address each of these issues by proposing a class of multinomial, closed-form models with zero or a finite number of shape parameters. The proposed class will be able to avoid the symmetry property without restrictions on the index, and it will include many of the existing models as special cases. We will also provide guidance on extending existing binary models to the multinomial setting and on creating new asymmetric choice models.

\section{A General Class of Asymmetric Models}
\label{sec:logit_type_models}
In this section, we present a class of discrete choice models that can avoid the symmetry property described in Section \ref{sec:intro} without imposing restrictions on the sign or magnitude of the index, $V_{ij}$, for any given alternative $j$. We will proceed as follows. Section \ref{sec:logit_type_formulation} will give the general formulation of our proposed class of models and show how this formulation can avoid the symmetry property described above. Section \ref{sec:logit_type_literature_relation} will then relate our models to existing literature. Next, in Section \ref{sec:binary_to_multinomial} we will demonstrate how our proposed class of models can be used to extend existing, asymmetric, closed-form models of binary choice to the multinomial setting. To do so, we will extend the clog-log model and the scobit model from the binary to the multinomial setting for the first time. Finally, in Section \ref{sec:deriving_binary_asym_models} we will propose and demonstrate one possible approach to deriving new asymmetric choice models when existing models are not adequate for one's needs. In doing so, we will derive two new asymmetric choice models, the ``uneven logit model'' and the ``asymmetric logit model.''

\subsection{General Formulation}
\label{sec:logit_type_formulation}
Our proposed class of models, described below, is appropriate for multinomial choice situations, has a closed-form probability equation, and only depends on a finite number of parameters. Moreover, we refer to our proposed model class as ``Logit-Type'' models because their choice probability functions share the same functional form as the MNL model: an exponential term divided by a sum of exponential terms. The choice probability function for our proposed ``logit-type''  models is:
\begin{equation}
\label{eq:logit_type_models}
\begin{aligned}
P \left( y_{ij} = 1 | \tau, \gamma, V_{i1}, V_{i2}, ..., V_{ik} \ \forall \left\lbrace j, k \right\rbrace \in C_i \right) &= \frac{\exp \left[ \tau _j + S \left( V_{ij}, \gamma _j \right) \right]}{ \sum _{\ell \in C_i} \exp \left[ \tau _{\ell} + S \left( V_{i \ell}, \gamma _{\ell} \right) \right]}  \\
&= \frac{\exp \left( S_{ij} \right)}{ \sum _{\ell \in C_i} \exp \left( S_{i \ell} \right)}\\
\textrm{where } \tau &= \textrm{a 1-dimensional vector of constants, with one value}\\
&\quad \ \textrm{for each alternative in the dataset.}\\
\gamma &= \textrm{a 2-dimensional matrix of shape parameters, with}\\
&\quad \ \textrm{one column for each alternative in the dataset.}\\
\tau _j &= \textrm{a constant associated with alternative $j$.}\\
\gamma _j &= \textrm{a column vector of shape parameters}\\
&\quad \ \textrm{associated with alternative $j$.} \\
S \left( \cdot, \cdot \right) &= \textrm{a closed-form, model-specific function of $V_{ij}$ and} \\
&\quad \ \textrm{$\gamma_j$. It is monotonically increasing in $V_{ij}$. As before,} \\
&\quad \ \textrm{if a model has no shape parameters, then we replace}\\
&\quad \ \textrm{$S \left( V_{ij}, \gamma _j \right)$ with $S \left( V_{ij} \right)$.}\\
\end{aligned}
\end{equation}

Note that unlike standard logit models, our class of logit-type models makes no assumptions regarding additive random utility functions. As shown by Mattsson et al. (\citeyear{mattsson_extreme_2014}), models of the form given in Equation \ref{eq:logit_type_models} are obtainable under an infinite number of random utility specifications, not all of which are additive. One example we have already mentioned is the case of multiplicative utilities \citep{castillo_closed_2008, fosgerau_discrete_2009} that are Weibull distributed. As shown by this example, it may be incorrect to interpret $S_{ij} = \tau _j + S \left( V_{ij}, \gamma _j \right)$ as a redefinition of the systematic portion of one's utility. Expressions such as $S_{ij}$ may arise solely due to the derivation of the choice probabilities, and they may not actually be present in the simplest expression of the random utility.

To see how our proposed model class can avoid the symmetry property described in the introduction, one can make the analogy between the logit-type models given in Equation \ref{eq:logit_type_models} and the MNL model given in Equation \ref{eq:mnl_formula}. Since the two models share the exact same functional form except for the replacement of $V_{ij}$ with $S_{ij}$ and since variable names do not influence mathematical properties, it follows that logit-type models are symmetric with respect to $S_{ij}$. This means that for logit-type models, equal-magnitude increases and decreases in the probability of choosing alternative $j$ (from an initial probability of 50\%) will result if and only if equal-magnitude increases and decreases in $S_{ij}$ are experienced. As a consequence, one will avoid the aforementioned symmetry property if and only if equal-magnitude increases and decreases in $V_{ij}$ (respectively) lead to unequal increases and decreases in $S_{ij}$. Formally, if 
\begin{equation}
S \left( V_{ij} + \varphi, \gamma _j \right) - S \left( V_{ij}, \gamma _j \right) \neq S \left( V_{ij}, \gamma _j \right) - S \left( V_{ij} - \varphi, \gamma _j \right), \forall \ \varphi \geq 0
\end{equation}
then the logit-type model given by Equation \ref{eq:logit_type_models} will be asymmetric with respect to $V_{ij}$.

\subsection{Relation to existing literature}
\label{sec:logit_type_literature_relation}
The logit-type models described above encapsulate and generalize many closed-form, finite-parameter, discrete choice models that exist in the literature\footnote{See Appendix B (Section \ref{appendix:lit_relations}) for a convenient table that explicitly shows how our logit-type models include previous models from the literature as special cases.}. For example, one can use Equation \ref{eq:logit_type_models} to denote the models described by Li (\citeyear{li_multinomial_2011}) that were based on assuming Weibull, Rayleigh, Type II Generalized Logistic, Pareto, or Exponential distributions for one's utilities. Except for the model based on the weibull distribution, $\tau_j = 0 \ \forall j$  and there are no shape parameters. When basing one's choice model on Weibull distributed utilities, we have $\gamma _j = \gamma^{ \star } \ \forall j$ where $\gamma^{\star}$ is the scale parameter of the distributions. The precise transformations, $S \left( \cdot \right)$, are provided in Table 2 of Li (\citeyear{li_multinomial_2011}) for each distribution mentioned above\footnote{See Appendix B (Section \ref{appendix:lit_relations}) for more details on how our proposed logit-type model are related to and different from the model of \citet{li_multinomial_2011} and \citet{das_generalized_2014}.}. Likewise the asymmetric, closed-form, multinomial choice model of Das and Mukhopadhyay (\citeyear{das_generalized_2014}) is a special case of the models given in Equation \ref{eq:logit_type_models}. Here, again, $\tau_j = 0 \ \forall j$. However, their model has two shape parameters for each alternative, so $\gamma$ has two rows, and $S \left( \cdot \right)$ is now given by their function $G \left( \cdot \right)$ \citep{das_generalized_2014}. Examples of binary models that are special cases of logit-type models will be given in Section \ref{sec:binary_to_multinomial} when we demonstrate how one can use Equation \ref{eq:logit_type_models} to extend existing binary models to the multinomial setting.

To be clear, not all closed-form, finite-parameter discrete choice models are special cases of logit-type models. An example of a closed-form, finite-parameter, multinomial model that is not a special case of a logit-type model is the ``exponomial choice model'' (also known as the ``negative exponential distribution'' model) \citep{daganzo_multinomial_1979, alptekinoglu_exponomial_2016}. This can be most easily seen by considering the fact that logit-type models do not depend on the order statistics (i.e. the rankings from lowest to highest) of the indices, $V_{ij}$. In contrast to this, the probabilities predicted by the exponomial choice model depend on both the magnitude and the order statistics of each $V_{ij}$. Furthermore, not all models with choice probabilities given by a ratio of an exponential term in the numerator divided by a sum of exponential terms in the denominator are logit-type models. Examples of this are the ``Random Regret Minimization'' and ``Relative Advantage Maximization'' models where each exponential term depends on variables related to all of the alternatives \citep{chorus_random_2014, leong_contrasts_2015}. In our logit-type models, each exponential term depends only on the attributes of one alternative (through $S\left(V_{ij}, \gamma_j\right)$).

While logit-type models generalize many closed-form, individual choice models that have been described in the literature, they are also a particular parametrization of the class of models described by Mattsson et al. (\citeyear{mattsson_extreme_2014}). Using the notation of Mattsson et al., we can show equivalence between the two model classes if we set $w_{j} = \exp \left[ \tau _j + S \left( V_{j}, \gamma _j \right) \right]$, where the index $i$ has been suppressed to match the notation used in the Mattsson et al. paper (which described the choice probabilities of a single individual). Viewing logit-type models through the lens of the Mattsson et al. paper is useful for two reasons. First, the Mattsson et al. paper provides a rigorous justification for the multinomial specifications of our logit-type models. Secondly, when thinking of further extensions to our work, the Mattsson et al. paper explains why we cannot automatically generalize our logit-type models to models that are analogous to the nested logit model. In particular, Mattsson et al. show that specifying $S \left( \cdot \right)$ is necessary but not sufficient for specifying models that can cope with dependence between one's random utilities. To account for this dependence (as with nested logit), one needs to also specify an ``aggregation function'' that dictates how the various random utilities are combined into a joint distribution. However, determination of such aggregation functions is an open question that we do not attempt to address because it is beyond the scope of this paper.

To recap, the logit-type models introduced in Section \ref{sec:logit_type_formulation} are both a generalization of many existing models and a special case of a wider class of models introduced by Mattsson et al. (\citeyear{mattsson_extreme_2014}). This position allows us to easily extend previously existing binary choice models to the multinomial setting, thereby making them more useful for transportation researchers. Simultaneously, this position allows us to rely on the theoretical justifications that Mattsson et al. provide for our entire class of models. The next subsection will focus on multinomial extensions of binary choice models in greater detail. Looking further ahead, we have noted that there are choice models that are either not part of the logit-type framework or are non-trivial extensions of logit-type models. These non-logit-type models are not considered in this paper, but they point to the need for more research to expand the types of models that avoid the symmetry property described in Section \ref{sec:intro}. This point will be returned to in Section \ref{sec:extensions} where we describe the possible future work that stems this paper.

\subsection{Extending Binary Models to the Multinomial Setting}
\label{sec:binary_to_multinomial}
As noted in the literature review (Section \ref{sec:lit_review}), there are many asymmetric, binary choice models, but these models have limited usefulness for transportation researchers because many choice contexts in transportation are inherently multinomial. In this subsection, we propose a technique for using our class of logit-type models given by Equation \ref{eq:logit_type_models} to create multinomial extensions of existing binary choice models. As a result, transportation scholars and practitioners will be better able to leverage the work that has already been done to create the asymmetric binary choice models that exist in the literature. First, we will describe our procedure, and then we will demonstrate it with two examples. In particular, we will create multinomial generalizations of the binary clog-log model\footnote{Note, the clog-log model has \textit{\textbf{not}} been chosen based on any arguments related to its predictive performance in previous studies. As one referee points out, the clog-log model does not always perform well as compared to its logit and probit counterparts. In fact, as noted in Section \ref{sec:application_discussion} where we discuss the results of this paper's empirical applications, the clog-log model does not perform well given our study's dataset. However, there are numerous documented cases where the clog-log model does perform well relative to the logit and probit model, for example \citet{spiegelhalter_bayesian_2002, presnell_ios_2004}. Accordingly, the clog-log model should not be disregarded a-priori. Nevertheless, we re-emphasize that the clog-log model is included here because it is one of the most commonly used asymmetric probability functions within the statistical literature, and may therefore serve as a more illuminating example as compared to less common asymmetric probability functions.} \citep{yates_use_1955} and the binary scobit model \citep{nagler_scobit:_1994}. These two models are chosen, in part, because the clog-log model is one of the oldest and most well-known asymmetric discrete choice models \citep{fisher_mathematical_1922, yates_use_1955, mccullagh_generalized_1989} and because the scobit model has been used in multiple disciplines such as political science \citep{nagler_scobit:_1994}, transportation \citep{zhang_scobit-based_2010, zhang_developing_2011, wu_analysis_2012}, and finance \citep{golet_symmetric_2014}. 

Overall, our procedure for using Equation \ref{eq:logit_type_models} to extend existing, closed-form, binary choice models to the multinomial setting is given in Table \ref{table:procedure_for_creating_multinomial_extensions}. The basic idea behind this procedure is that \textbf{\textit{if}} we can express an existing, binary, choice probability function as $\frac{ \exp \left( S_{ij} \right) }{ \sum _{ \ell = 1 } ^2 \exp \left( S_{i \ell} \right) }$, then the work of Mattsson et al. (\citeyear{mattsson_extreme_2014}) rigorously shows that there are an infinite number of random utility formulations that could have lead to the given binary choice probabilities. Moreover, Mattsson et al. showed that the same utility formulations, with more alternatives, would lead to choice probabilities of the form given in Equation \ref{eq:logit_type_models}. The extension from a sum of two exponential terms in the denominator of the binary model, to a sum of three or more exponential terms is thereby well-founded.

Following Table \ref{table:procedure_for_creating_multinomial_extensions}, we demonstrate our procedure with two examples, deriving multinomial versions of the clog-log and scobit models for the first time. For an example of using this procedure to generalize the binary logit model to the MNL, see Appendix C, where we perform the extension to the multinomial setting in the context of providing a new derivation of the binary logit and the MNL models.

\begin{table}[h!]
\centering
\caption{Procedure for Creating Multinomial Extensions of Binary Choice Models}
\label{table:procedure_for_creating_multinomial_extensions}
\begin{tabular}{| l K{0.915\linewidth} |}
\hline

1. & Determine if one's existing binary choice model is given in terms of both $V_{i1}$ and $V_{i2}$ or if it is only given in terms of $V_{i1}$.  \\
{} & {} \\

2. &  If one's binary choice model is only given in terms of $V_{i1}$: \\
\hphantom{2.}(a) & Assume $V_{i2} = x_{i2} = 0$. \\
\hphantom{2.}(b) & Solve for $S \left( V_{i1}, \gamma _1 \right)$ to identify the functional form of $S \left( \cdot \right)$. \\
\hphantom{2.}(c) & Calculate $S \left( V_{i2}, \gamma _2 \mid V_{i2} = 0 \right)$. \\
\hphantom{2.}(d) & Use $S \left( V_{i1}, \gamma _1 \right)$ and $S \left( V_{i2}, \gamma _2 \mid V_{i2} = 0 \right)$ to determine any restrictions on the values of $\tau$ and $\gamma$ that need to be made to establish the binary choice model as a special case of the logit-type models. \\
{} & {} \\

3. & If one's binary choice model is given in terms of both $V_{i1}$ and $V_{i2}$: \\ 
\hphantom{3.}(a) & Express one's existing choice model as a fraction with one term in the numerator and a sum of terms in the denominator. \\
\hphantom{3.}(b) & Ensure that each term in the numerator and denominator contains only one index $V_{ij}$. \\
\hphantom{3.}(c) & Directly solve for $S_{ij}$, for all alternatives $j$. \\
\hphantom{3.}(d) & Determine any restrictions on the values of $\tau$ and $\gamma$ that need to be made to establish the binary choice model as a special case of the logit-type models. \\
{} & {} \\

4. & Relax all restrictions from the previous two steps to generalize the binary model and to create multinomial versions of the model. \\

\hline 
\end{tabular}
\end{table}

\subsubsection{Example 1: Deriving the Multinomial Clog-log Model}
\label{sec:deriving_multinomial_cloglog}

The binary clog-log model \citep{fisher_mathematical_1922, yates_use_1955, mccullagh_generalized_1989} was introduced within the field of statistics, where there are usually no explanatory variables that vary with one's alternatives. The choice probability function of the binary clog-log model is commonly written as $P_{\textrm{clog-log}} \left( y_{ij} = 1 \mid V_{i1} \right) = 1 - \exp \left( - \mathrm{e}^{V_{i1}} \right)$, and the function is plotted in both Figures \ref{fig:symm_logit_probit} and \ref{fig:asym_models}. As statisticians typically do when there are no explanatory variables that vary with one's alternatives, the probability of the outcome of interest ($y_{i1} = 1$) is spoken of as being only a function of $V_{i1} = x_{i1} \beta$, without any regard for $V_{i2} = x_{i2} \beta$. Often, $V_{i2}$ is not even defined. For instance, when statisticians speak of binary logit models, they usually write $P \left( y_{i1} = 1 \mid x_{i1}, \beta \right) = \left[ 1 + \exp \left( - x_{i1} \beta \right) \right]^{-1}$ whereas econometricians and transportation researchers would equivalently say $P \left( y_{i1} = 1 \mid x_{i1}, x_{i2}, \beta \right) = \left[ 1 + \exp \left( V_{i2} - V_{i1} \right) \right]^{-1} = \left[ 1 + \exp \left( \left\lbrace x_{i2} - x_{i1} \right\rbrace \beta \right) \right]^{-1}$. Clearly, the unstated assumption is that $x_{i2} = 0$. Whenever binary discrete choice models fail to define $V_{i2}$, we will adopt the convention\footnote{Note, if one's vector of explanatory variables contains alternative specific variables, such as costs for each alternative, then one's binary choice model is likely to be implicitly defined in terms of $V_{i1}$ and $V_{i2}$. If one's choice model is given in terms of both $V_{i1}$ and $V_{i2}$, yet Table \ref{table:procedure_for_creating_multinomial_extensions} steps 3a and 3b cannot be performed, then it is likely that one's model is not expressible as a logit-type model, and it is likely that one cannot extend one's model to the multinomial setting using the procedures given in Table \ref{table:procedure_for_creating_multinomial_extensions}. This precludes the use of the procedures in Table \ref{table:procedure_for_creating_multinomial_extensions} for a clog-log model that is expressed as  $P_{\textrm{clog-log}} \left( y_{ij} = 1 \mid x_{i1}, x_{i2}, \beta \right) = 1 - \exp \left( - \mathrm{e}^{V_{i1} - V_{i2}} \right)$ or a scobit model that is expressed as $P_{\textrm{scobit}} \left( y_{ij} = 1 \mid x_{i1}, x_{i2}, \beta,  \gamma _1 \right) = \left( 1 + \exp\left[V_{i2} - V_{i1} \right] \right)^ {- \gamma _1}$. To the best of our knowledge, alternative specific explanatory variables have never been used with the clog-log and scobit models. Our paper therefore covers the most common use cases for these models. Thanks to an anonymous reviewer for prompting us to clarify this point.} that $V_{i2} = x_{i2} \beta = 0$. With this in mind, we can express the binary clog-log model as a special case of our logit-type models as follows:

\begin{equation}
\label{eq:clog_to_logit_part1}
\begin{aligned}
P_{\textrm{clog-log}} \left( y_{ij} = 1 \mid x_{i1}, x_{i2} = 0, \beta \right) &= 1 - \exp \left( - \mathrm{e}^{V_{i1}} \right) \qquad &\textrm{\textit{Step 2a.}}\\ 
1 - \exp \left( - \mathrm{e}^{V_{i1}} \right) &\equiv \frac{\exp \left( S_{i1} \right)}{\sum _{\ell = 1} ^2 \exp \left( S_{i \ell} \right)}  \qquad &\textrm{\textit{Step 2b.}}\\
\frac{P_{\textrm{clog-log}}}{1 - P_{\textrm{clog-log}}} = \frac{1 - \exp \left( - \mathrm{e}^{V_{i1}} \right)}{\exp \left( - \mathrm{e}^{V_{i1}} \right)} &= \frac{\exp \left( S_{i1} \right)}{\exp \left( S_{i2} \right)} \\
= \exp \left( \mathrm{e}^{V_{i1}} \right) - 1 &= \exp \left( S_{i1} - S_{i2} \right)\\
\ln \left[ \exp \left( \mathrm{e}^{V_{i1}} \right) - 1 \right] &= S_{i1} - S_{i2} \\
\ln \left[ \exp \left( \mathrm{e}^{V_{i1}} \right) - 1 \right]  &= \tau_1 + S \left( V_{i1} \right) - \tau_2 - S \left( V_{i2} \right)
\end{aligned}
\end{equation}
On the last line of the right hand side of Equation \ref{eq:clog_to_logit_part1}, only $S \left( V_{i1} \right)$ involves $V_{i1}$. This means that $S \left( V_{i1} \right) = \ln \left[ \exp \left( \mathrm{e}^{V_{i1}} \right) - 1 \right]$, and more generally, $S \left( V_{ij} \right) = \ln \left[ \exp \left( \mathrm{e}^{V_{ij}} \right) - 1 \right]$. This fact can be derived as follows. First, note that $S \left( V_{i1} \right)$ does not contain any arbitrary constants, as these can be thought of as part of $\tau_1$. Next, let $h \left(V_{i1} \right) = \ln \left[ \exp \left( \mathrm{e}^{V_{ij}} \right) - 1 \right]$. Then,
\begin{equation*}
\begin{aligned}
h \left(V_{i1} \right) &= \tau_1 + S \left( V_{i1} \right) - \tau_2 - S \left( V_{i2} \right) \\
\frac{\partial \left( h \left( V_{i1} \right) \right)}{\partial V_{i1}} &= \frac{\partial \left[ \tau_1 + S \left( V_{i1} \right) - \tau_2 - S \left( V_{i2} \right) \right]}{\partial V_{i1}} \\
\frac{\partial h \left( V_{i1} \right) }{\partial V_{i1}} &= \frac{\partial S \left( V_{i1} \right)}{\partial V_{i1}} \\
\partial h \left( V_{i1} \right) &= \partial S \left( V_{i1} \right) \\
\int \partial h \left( V_{i1} \right) &= \int \partial S \left( V_{i1} \right) \\
h \left( V_{i1} \right) &= S \left( V_{i1} \right) + A \  &\textrm{where A is a constant of integration} \\
h \left( V_{i1} \right) &= S \left( V_{i1} \right) \ &\textrm{because $S \left( V_{i1} \right)$ contains no arbitrary constants} \\
\ln \left[ \exp \left( \mathrm{e}^{V_{i1}} \right) - 1 \right] &= S \left( V_{i1} \right)
\end{aligned}
\end{equation*}

With this specification of $S \left( \cdot \right)$, we can further simplify Equation \ref{eq:clog_to_logit_part1} as follows:

\begin{equation}
\label{eq:clog_to_logit_part2}
\begin{aligned}
\ln \left[ \exp \left( \mathrm{e}^{V_{i1}} \right) - 1 \right]  &= \tau_1 + S \left( V_{i1} \right) - \tau_2 - S \left( V_{i2} \right)\\
S \left( V_{i1} \right) &=  \tau_1 + S \left( V_{i1} \right) - \tau_2 - S \left( V_{i2} \right) \\
0 &=  \tau_1 - \tau_2 - S \left( V_{i2} \right) \\
0 &= \tau_1 - \tau_2 - S \left( 0 \right) \qquad &\textrm{\textit{Step 2c.}}\\
0 &= \tau_1 - \tau_2 - \ln \left[ \exp \left( \mathrm{e}^{0} \right) - 1 \right] \\
0 &= \tau_1 - \tau_2 - \ln \left[ \mathrm{e} - 1 \right] \\
\ln \left[ \mathrm{e} - 1 \right] &= \tau_1 - \tau_2  \qquad &\textrm{\textit{Step 2d.}}
\end{aligned}
\end{equation}
From Equation \ref{eq:clog_to_logit_part2}, we have two unknowns $\tau_1$ and $\tau_2$, and one equation. Without loss of generality, we can therefore set $\tau_2 = 0$ and $\tau_1 = \ln \left[ \mathrm{e} - 1 \right]$. With these restrictions, we have shown that the binary clog-log model is a special case of the logit type models given by Equation \ref{eq:logit_type_models}, where there are no shape parameters $\gamma$, and where $S \left( V_{ij} \right) = \ln \left[ \exp \left( \mathrm{e}^{V_{ij}} \right) - 1 \right]$, $\tau_1 = \ln \left[ \mathrm{e} - 1 \right]$, and $\tau_2 = 0$.

From these results, we can form a ``conditional clog-log model'' that parallels the ``conditional logit model'' \citep{mcfadden_conditional_1972} and is immediately made useful to transportation researchers and econometricians by allowing explanatory variables that differ across alternatives. To do so, we merely remove the restriction that $x_{i2} = 0 \ \forall i$, and we remove the constraint that $ \tau_1 - \tau_2 = \ln \left[ \mathrm{e} - 1 \right]$. Of course, as with alternative specific constants in general, only the difference $\tau_2 - \tau_1$ is identified, so one of the two constants should be constrained. This ``conditional clog-log model'' can easily be extended to the multinomial setting in an analogous fashion to the multinomial logit model. Specifically, the multinomial clog-log model is given by Equation \ref{eq:logit_type_models}, where $S \left( V_{ij}, \gamma_j \right) = S \left( V_{ij} \right) =  \ln \left[ \exp \left( \mathrm{e}^{V_{ij}} \right) - 1 \right]$ as derived above, and where as usual, one of the $\tau_j$'s is constrained to zero for identification purposes. For convenience, the probability equation of the multinomial clog-log model is displayed below.

\begin{equation*}
P_{\textrm{clog-log}} \left( y_{ij} = 1 \mid \tau, V_{i1}, V_{i2}, ..., V_{ik} \ \forall \left\lbrace j, k \right\rbrace \in C_i \right) = \frac{\exp \left( \tau _j + \ln \left[ \exp \left( \mathrm{e}^{V_{ij}} \right) - 1 \right] \right)}{ \sum _{\ell \in C_i} \exp \left( \tau _{\ell} + \ln \left[ \exp \left( \mathrm{e}^{V_{i \ell}} \right) - 1 \right] \right)}
\end{equation*}

\subsubsection{Example 2: Deriving the Multinomial Scobit Model}
\label{sec:deriving_multinomial_scobit}

The multinomial scobit model is derived from the binary scobit model (see Figure \ref{fig:asym_models}) using the same process as with multinomial clog-log model. Given that the binary scobit model \citep{nagler_scobit:_1994} is defined only in terms of $V_{i1}$, we assume $V_{i2} = x_{i2} = 0$. From here we write,

\begin{equation}
\label{eq:scobit_to_logit_part1}
\begin{aligned}
P_{\textrm{scobit}} \left( y_{ij} = 1 \mid x_{i1}, x_{i2} = 0, \beta, \gamma _1 \right) &= \frac{1}{\left( 1 + \mathrm{e}^{- V_{i1}} \right)^ {\gamma _1}}, \quad \gamma_1 \in \left( 0, \infty \right) \qquad &\textrm{\textit{Step 2a.}} \\ 
\frac{1}{\left( 1 + \mathrm{e}^{- V_{i1}} \right)^ {\gamma _1}} &\equiv \frac{\exp \left( S_{i1} \right)}{\sum _{\ell = 1} ^2 \exp \left( S_{i \ell} \right)} \qquad &\textrm{\textit{Step 2b.}}\\
\left( 1 + \mathrm{e}^{- V_{i1}} \right)^ {- \gamma _1} &= \left[ 1 + \exp \left( S_{i2} - S_{i1} \right) \right]^{-1} \\
\left( 1 + \mathrm{e}^{- V_{i1}} \right)^ {\gamma _1} &= 1 + \exp \left( S_{i2} - S_{i1} \right) \\
\left( 1 + \mathrm{e}^{- V_{i1}} \right)^ {\gamma _1} - 1 &=  \exp \left( S_{i2} - S_{i1} \right) \\
\ln \left[ \left( 1 + \mathrm{e}^{- V_{i1}} \right)^ {\gamma _1} - 1 \right] &= S_{i2} - S_{i1} \\
\ln \left[ \left( 1 + \mathrm{e}^{- V_{i1}} \right)^ {\gamma _1} - 1 \right] &= \tau_2 + S \left( V_{i2}, \gamma_2 \right) - \tau_1 - S \left( V_{i1}, \gamma_1 \right), \quad \gamma_2 \in \left( 0, \infty \right)
\end{aligned}
\end{equation}
As before, since $S \left( V_{i1} , \gamma_1\right)$ is the only term on the right hand side of Equation \ref{eq:scobit_to_logit_part1} that contains $V_{i1}$, we can determine that $S \left( V_{i1}, \gamma_1 \right) = - \ln \left[ \left( 1 + \mathrm{e}^{- V_{i1}} \right)^ {\gamma _1} - 1 \right] $, and that even more generally, $S \left( V_{ij}, \gamma_j \right) = - \ln \left[ \left( 1 + \mathrm{e}^{- V_{ij}} \right)^ {\gamma _j} - 1 \right]$. Substituting these terms into Equation \ref{eq:scobit_to_logit_part1}, we can further simplify that equation to:

\begin{equation}
\label{eq:scobit_to_logit_part2}
\begin{aligned}
\ln \left[ \left( 1 + \mathrm{e}^{- V_{i1}} \right)^ {\gamma _1} - 1 \right] &= \tau_2 + S \left( V_{i2}, \gamma_2 \right) - \tau_1 - S \left( V_{i1}, \gamma_1 \right)\\
- S \left( V_{i1}, \gamma_1 \right) &= \tau_2 + S \left( V_{i2}, \gamma_2 \right) - \tau_1 - S \left( V_{i1}, \gamma_1 \right) \\
0 &= \tau_2 + S \left( V_{i2}, \gamma_2 \right) - \tau_1 \\
0 &= \tau_2 - \ln \left[ \left( 1 + \mathrm{e}^{- V_{i2}} \right)^ {\gamma _2} - 1 \right] - \tau_1 \\
0 &= \tau_2 - \ln \left[ \left( 1 + \mathrm{e}^{0} \right)^ {\gamma _2} - 1 \right] - \tau_1  \qquad &\textrm{\textit{Step 2c.}}\\
\tau_1 - \tau_2 &= \ln \left[ 2^{\gamma _2} - 1 \right] \qquad &\textrm{\textit{Step 2d.}}
\end{aligned}
\end{equation}
Here, we have more unknowns than equations, so some of the parameters are not identified and must be constrained. If we set $\gamma _2 = 1$, then this means $\tau_1 = \tau_2$, and without loss of generality, we can assume $\tau_1 = \tau_2 = 0$. With these constraints, we have shown that the binary scobit model is a special case of the logit-type models given by Equation \ref{eq:logit_type_models}.

As with the binary clog-log model, the binary scobit model can be immediately generalized to a ``conditional scobit model'' that allows for explanatory variables that differ across alternatives. The ``conditional scobit model'' is derived by removing the constraints $\gamma _2 = 1$, $x_{i2} = 0$, and $\tau_1 = \tau_2 = 0$. As usual, one of the alternative specific constants, $\tau_1$ or $\tau_2$, must still be constrained for identification purposes.

Finally, as with the multinomial clog-log model, the generalization of the ``conditional scobit model'' to the multinomial setting is immediate. The multinomial scobit model is given by Equation \ref{eq:logit_type_models}, where $S \left( V_{ij}, \gamma_j \right) = - \ln \left[ \left( 1 + \mathrm{e}^{- V_{ij}} \right)^ {\gamma _j} - 1 \right]$ and where $\gamma _j$ is a scalar, for each alternative $j$, that is to be estimated along with $\beta$ and all but one of the $\tau _j$'s (for identifiability). For convenience, the probability formula for the multinomial scobit model is displayed below.

\begin{equation*}
P_{\textrm{scobit}} \left( y_{ij} = 1 \mid \tau, \gamma, V_{i1}, V_{i2}, ..., V_{ik} \ \forall \left\lbrace j, k \right\rbrace \in C_i \right) = \frac{\exp \left( \tau _j - \ln \left[ \left( 1 + \mathrm{e}^{- V_{ij}} \right)^ {\gamma _j} - 1 \right] \right)}{ \sum _{\ell \in C_i} \exp \left( \tau _{\ell} - \ln \left[ \left( 1 + \mathrm{e}^{- V_{i \ell}} \right)^ {\gamma _{\ell}} - 1 \right] \right)}
\end{equation*}

\subsection{Creating New Asymmetric Choice Models}
\label{sec:deriving_binary_asym_models}
In Section \ref{sec:binary_to_multinomial}, we showed how one can extend existing, binary choice models to the multinomial setting. In this section, we will present our method for creating new binary choice models. Note that our proposed process is general enough to create both new asymmetric and new symmetric choice models. Together, Section \ref{sec:binary_to_multinomial} and Section \ref{sec:deriving_binary_asym_models} provide a way to create new multinomial choice models. In this paper, however, we will focus on the creation of new asymmetric, multinomial choice models. The rest of this subsection will proceed as follows. First, we will briefly review traditional methods in transportation for creating new choice models, and why we think such methods are not easy to use. Next, we will present an alternative approach for creating new binary choice models---an approach that does not begin by specifying the distribution of error terms in one's utility functions. We will then review the key concepts necessary to understand this approach, and finally, we will present two examples where we demonstrate the procedure by creating new, asymmetric, binary choice models and extending them to the multinomial setting.

As noted by Ben-Akiva and Lerman, ``varying the assumptions about the distributions of [one's utilities] [...] leads to different choice models'' \citep[p.65]{ben-akiva_discrete_1985}. This approach of first specifying the distribution of one's utilities, and then deriving one's choice probabilities, is commonly used in transportation. For instance, it is used by the transportation researchers cited above such as Castillo et al. (\citeyear{castillo_closed_2008}), Fosgerau and Bierlaire (\citeyear{fosgerau_discrete_2009}), Li (\citeyear{li_multinomial_2011}), and Mattsson et al. (\citeyear{mattsson_extreme_2014}). While clearly a viable approach, discrete choice analysts have acknowledged that ``it will often be difficult to make strong statements about the overall distribution of [one's utilities]'' \citep[p.66]{ben-akiva_discrete_1985}. To sidestep these difficulties Daniel McFadden (emphasis is his own) wrote that: 
\begin{quotation}
``In practice, it is difficult to define joint distributions [of one's utilities] which allow the computation of econometrically useful formulas for the [selection probabilities]. An alternative approach is to specify formulas for the selection probabilities and then examine the question of whether these formulas could be obtained [...] from \textit{some} distribution of utility-maximizing consumers'' \citep[p.108]{mcfadden_conditional_1972}.
\end{quotation}
This approach of directly specifying probability formulas is the one that we will take. From the work of Mattsson et al. (\citeyear{mattsson_extreme_2014}), we know that any choice model of the form given in Equation \ref{eq:logit_type_models} can be generated from an infinite number of joint distributions of one's utilities. Moreover, we know that the newly derived logit-type models will be ``well-behaved.'' To be specific, because logit type models have an exponential term for their numerator and a sum of exponential terms for their denominator, where the sum includes the numerator, logit-type models will always return probabilities between zero and one. Also, because $x_{ij}$ only appears in $S_{ij}$ and because $S_{ij}$ was defined as being a monotonically increasing function of $V_{ij} = x_{ij} \beta$, interpreting whether the probability of choosing alternative $j$ increases or decreases when we increase a variable in $x_{ij}$ remains as easy as it was with the standard MNL model. Often, such interpretation consists of just knowing the sign on the index-coefficient ($\beta$) of the variable of interest. Given these beneficial properties, we can generate new logit-type models simply by specifying $S \left( \cdot \right)$.

To specify the $S \left( \cdot \right)$ function in one's logit-type models, we created the three-step procedure\footnote{Note that our procedure was motivated by computer scientists who made use of asymmetric loss functions (defined in the coming paragraphs) when dealing with class-imbalanced datasets. When investigating this use of asymmetric loss functions, we came across literature that noted the fact that loss functions are related to specific choice probability functions. This discovery lead us to think that a useful way of deriving choice models, given the literature on choosing or designing loss functions, would be to first choose a desired loss function and then derive its related choice probability function.} shown in Table \ref{table:procedure_for_creating_new_models}. Note that this procedure will make use of potentially unfamiliar terms and concepts such as ``binary loss functions,'' ``asymmetric loss functions,'' ``properties of loss functions,'' and ``related, binary, choice probability functions.'' However, all of these terms will be explained and made more precise in the following paragraphs. After these explanations, we will demonstrate our procedure. First, we will use the process in Table \ref{table:procedure_for_creating_new_models} to re-derive the familiar MNL model. Then we will further demonstrate the procedure by creating two new, closed-form, asymmetric, choice probability functions.

\begin{table}[h!]
\centering
\caption{Procedure for Creating New Multinomial Choice Models}
\label{table:procedure_for_creating_new_models}
\begin{tabular}{| l K{0.915\linewidth} |}
\hline

1. & Choose a \textit{binary loss function} with properties that are desirable for one's study. If an asymmetric choice model is desired, then be sure to choose an \textit{asymmetric loss function}. \\
{} & {} \\

2. &  Derive the \textit{related, binary, choice probability function} for one's chosen loss function. \\
{} & {} \\

3. & Use the procedure detailed in Section \ref{sec:binary_to_multinomial} to convert one's derived choice probability function to a logit-type model. In the process, one will have determined $S \left( \cdot \right)$ and created a new multinomial choice model. \\ 

\hline 
\end{tabular}
\end{table}

Given that the first step in our proposed procedure is to choose a binary loss function with properties that are desirable for one's study, we will begin by defining loss functions, and then we will explain what is meant by properties of the loss function. Loss functions are functions that measure the quality of one's predictions, and binary loss functions measure the quality of one's predictions when one's observed, dependent variable takes on one of two possible values. Overall, there are two types of binary loss functions: ``class probability estimation (CPE) loss functions'' and ``composite loss functions'' \citep{reid_composite_2010}. For the purposes of Step 1 of our procedure, either of the two types of loss functions may be chosen. However, the two types of loss functions lead to differences in how the related choice probability functions are derived in Step 2 of our procedure. As a result, we will briefly describe both types of losses in the next paragraph. Additionally, we will make connections with concepts that most readers will be familiar with by showing the CPE and composite loss functions that are related to the binary logit model. 

We will start with CPE loss functions. CPE losses take an observed outcome and the predicted probability of that outcome occurring as arguments, and they output a penalty (i.e. a non-negative value) for discrepancies between the observation and prediction \citep{reid_composite_2010}. Typically, the returned penalty increases as the magnitude of the discrepancy increases. For example, the CPE loss function that is related to the binary logit model is the negative log-likelihood. This CPE loss is given by
\begin{equation}
\label{eq:neg_log_likelihood}
\begin{aligned}
\textrm{Negative Log-Likelihood}\left(y_{i1},  P \left( y_{i1} = 1 \mid V_{i1}, V_{i2} \right) \right) &= 1_{\left\lbrace y_{i1} = 1 \right\rbrace} \left( - \ln \left[ P \left( y_{i1} = 1 \mid V_{i1}, V_{i2} \right) \right] \right) +\\
&\quad \  1_{\left\lbrace y_{i1} = 0 \right\rbrace} \left( - \ln \left[ 1 - P \left( y_{i1} = 1 \mid V_{i1}, V_{i2} \right) \right] \right)
\end{aligned}
\end{equation}
where $1_{\left\lbrace r \right\rbrace}$ is an indicator function that equals 1 if $r$ is true and 0 otherwise.

Moving to composite losses, we noted in Section \ref{sec:deriving_multinomial_cloglog} that statisticians and computer scientists often speak of $P \left( y_{i1} = 1 \right)$ as being only a function of $V_{i1} = x_{i1} \beta$, without any regard for $V_{i2} = x_{i2} \beta$. In such settings, where it is often implicitly the case that $x_{i2} = 0$, one can speak of ``composite loss functions,'' that are simply functions of $V_{i1}$. Formally, composite loss functions are CPE loss functions composed of the choice probability function $P \left( y_{i1} = 1 \mid V_{i1} \right)$ \citep{reid_composite_2010}. As an example, consider the composite loss function that is related to the binary logit model---the log-loss. This loss function is derived by composing the negative log-likelihood given in Equation \ref{eq:neg_log_likelihood} with the choice probability function, $P \left( y_{i1} = 1 \mid V_{i1} \right) = \left[ 1 + \exp \left( -V_{i1} \right) \right]^{-1}$. We will omit the algebra used to simplify the composition, but the log-loss is given by the following formula:
\begin{equation}
\label{eq:log_loss}
\begin{aligned}
\textrm{Log-Loss}\left(y_{i1},  V_{i1} \right) &= 1_{\left\lbrace y_{i1} = 1 \right\rbrace} \ln \left(1 + \mathrm{e}^{-V_{i1}} \right) + 1_{\left\lbrace y_{i1} = 0 \right\rbrace} \ln \left(1 + \mathrm{e}^{ V_{i1} } \right)
\end{aligned}
\end{equation}

Given the formulation of composite losses, these functions differ from CPE loss functions only in their arguments. While both losses return a penalty for the discrepancy between one's observed outcome and the predicted probability of that outcome occurring, composite loss functions take the observed outcome and  $V_{i1}$ (as opposed to $P \left( y_{i1} = 1 \mid V_{i1} \right)$) as arguments. Note that CPE loss functions are defined for arbitrary choice probability functions, including those of the form $P \left( y_{i1} = 1 \mid V_{i1}, V_{i2} \right)$, whereas composite loss functions are only defined for choice probability functions of the form $P \left( y_{i1} = 1 \mid V_{i1} \right)$. This is analogous to the situation described in Section \ref{sec:binary_to_multinomial} where one's choice probability function could depend only on $V_{i1}$ or on both  $V_{i1}$ and  $V_{i2}$. As in Table \ref{table:procedure_for_creating_multinomial_extensions}, different steps are taken based on the situation we are in.

Now, beyond merely choosing a loss function, step 1 of our procedure requires choosing a loss function based on its properties. To place such properties in context, we emphasize that for our purposes, the most important use of loss functions is as a tool for parameter estimation. In an optimization setting, loss functions are used in statistics and computer science to estimate parameters of interest, such as the $\beta$'s in one's choice model \citep{gneiting_strictly_2007, dawid_geometry_2006}. The idea is that one chooses the set of parameters that minimizes the total loss (i.e. the sum of the loss for each observation), given one's dataset. In a parameter estimation setting, each loss function has properties that impact the estimation process and results. One important property is whether or not a loss function is symmetric. Symmetric, binary loss functions output equal-magnitude penalties for equal magnitude discrepancies, regardless of the observed outcome. For example, imagine we are analyzing the losses incurred on two observations: observations 1 and 2. Observation 1 is associated with outcome 1, and observation 2 is associated with outcome 2. For both observations, we predicted a 30\% probability of that observation being associated with its actual outcome. A symmetric loss function would assign the same penalty to our predictions for both observation 1 and observation 2. In contrast, an asymmetric loss function would assign different penalties to observation 1 and observation 2 because asymmetric loss functions unequally penalize each outcomes' predicted probabilities.

Aside from symmetry, loss functions have other properties that impact one's parameter estimates. For example, one might consider whether one's loss function is strictly proper (i.e. the loss is Fisher consistent and increasing discrepancies \textit{always} lead to increasing penalties) \citep{buja_loss_2005, reid_composite_2010}; robust against outliers \citep{pregibon_resistant_1982, carroll_robustness_1993, bianco_robust_1996}; or sparsity-inducing (in terms of identifying ``irrelevant'' predictors'', i.e. setting their $\beta$ coefficient to zero) \citep{kyung_penalized_2010, bach_optimization_2012, xu_sparse_2012}. In general, there are numerous properties that might be of interest. As such, it is beyond the scope of this paper to (1) comprehensively review and describe these properties or (2) instruct readers on how to design their loss functions with respect to these various properties. Interested readers seeking guidance may refer to works such as \citet{hennig_thoughts_2007} or \citet{merkle_choosing_2013}. Our main point is that loss functions have properties, that analysts can choose the most desirable mix of properties for their research needs, and that once an analyst has designed or found a loss function with the appropriate properties for their study, a related choice probability function can be derived from the chosen loss function. The next paragraph will describe precisely what is meant by the term ``related choice probability function'' and how one can derive it.

In general, one can derive unique choice probability functions from both composite loss functions and strictly proper CPE loss functions. In the case of CPE loss functions, the related choice probability function is such that when minimizing one's total loss, one is guaranteed to have an optimization problem that is convex in one's $\beta$'s \citep{buja_loss_2005, reid_composite_2010}. In the case of composite loss functions, the related choice probability function is the one that must have been used to derive the composite loss \citep{reid_composite_2010}. In each case, the derivation of the related choice probability functions uses what are known as the partial losses for a binary loss function. The partial losses are simply the functions used to supply the penalty for predictions on each of the two possible discrete outcomes \citep{buja_loss_2005, reid_composite_2010}. Formally, a given binary loss function $L \left( y_{i1}, \cdot \right)$ can be written as $L \left( y_{i1}, \cdot \right) = 1_{\left\lbrace y_{i1} = 1 \right\rbrace} L_1 \left( \cdot \right) + 1_{\left\lbrace y_{i1} = 0 \right\rbrace} L_2 \left( \cdot \right)$. The second argument of $L \left( y_{i1}, \cdot \right)$ depends on whether or not we are using a CPE loss function or a composite loss function. In either case, $L_1 \left( \cdot \right)$ and $L_2 \left( \cdot \right)$ are known as the partial losses for $L \left( y_{i1}, \cdot \right)$. $L_1$ determines the penalty if the observation is associated with outcome 1, and $L_2$ determines the penalty if the observation is associated with outcome 2. To derive the related choice probability functions we will start with the simpler derivation, the one for composite loss functions. It has been proven that in order for a binary composite loss function with differentiable partial losses to have been created by the composition of a CPE loss function and a choice probability function, the choice probability function must satisfy the following criteria \citep[Eq. 11]{reid_composite_2010}:
\begin{equation}
\label{eq:composite_loss_to_binary_prob}
P \left( y_{i1} = 1 \mid V_{i1} \right) = \frac{ L_{2}' \left( V_{i1} \right) }{ L_{2}' \left( V_{i1} \right) - L_{1}' \left( V_{i1} \right) }
\end{equation}
We will use this equation directly in order to derive the related choice probability function for composite losses. For strictly proper CPE loss functions, if one makes the usual assumption from statistics and computer science that $V_{i2} = 0$, then there exists a canonical choice probability function that can be derived by solving the following differential equation\footnote{Our derivation of this formula is given in Appendix A (see Section \ref{appendix:proofs}).} for $P \left( y_{i1} = 1 \mid V_{i1}, V_{i2} = 0 \right)$:
\begin{equation}
\label{eq:cpe_loss_to_binary_prob}
\frac{d \left[ L_{2} \left( \hat{p} \left( V_{i1} \right) \right) \right]}{d \hat{p} \left( V_{i1} \right)} = \frac{\hat{p} \left( V_{i1} \right) }{ \hat{p}' \left( V_{i1} \right) } \quad \textrm{where } \hat{p} \left( V_{i1} \right) = P \left( y_{i1} = 1 \mid V_{i1}, V_{i2} = 0 \right)
\end{equation}
In the following examples, we will show how both of these equations can be used in our proposed procedure for generating new multinomial choice models. In particular, we will use our proposed procedure to create two new, asymmetric, closed-form, choice probability functions. We will create the \textit{uneven logit model} from a composite loss function and then create the \textit{asymmetric logit model} using a CPE loss function. To see these procedures used in a setting that is likely to be more familiar to discrete choice modelers, see Appendix C (Section \ref{appendix:deriving_mnl}). There, we derive the binary logit and MNL models from their related CPE loss (i.e. the negative log-likelihood) and related composite loss (i.e. the log-loss).

\subsubsection{Example 3: Creating the Uneven Logit Model}
\label{sec:deriving_uneven_logit}
In ``Calibrated asymmetric surrogate losses'' \citep{scott_calibrated_2012}, Scott provides a way of creating asymmetric, composite loss functions from symmetric ones. Scott's main goal was to create loss functions that performed optimally under different costs for wrong classification predictions (i.e. binary predictions as opposed to probability predictions). Since altering misclassification costs is one technique used to deal with class imbalance in computer science, we decided to see whether the choice probability functions derived from Scott's asymmetric composite losses would be useful for making probability predictions under class imbalance. 

To begin, we used the procedures in Scott's paper to derive the following ``uneven log-loss'':
\begin{equation}
\label{eq:uneven_log_loss}
\begin{aligned}
\textrm{Uneven log-loss} &= 1_{\left\lbrace y_{i1} = 1 \right\rbrace} L_1 \left( V_{i1} \right) + 1_{\left\lbrace y_{i1} = 0 \right\rbrace} L_2 \left( V_{i1} \right)\\
&= 1_{\left\lbrace y_{i1} = 1 \right\rbrace} \ln \left(1 + \mathrm{e}^{-V_{i1}} \right) + 1_{\left\lbrace y_{i1} = 0 \right\rbrace} \frac{1}{\gamma_1} \ln \left(1 + \mathrm{e}^{\gamma_1 V_{i1} } \right), \quad \gamma_1 >  0
\end{aligned}
\end{equation}
We then derive the related choice probability function as follows:
\begin{equation}
\label{eq:binary_uneven_logit}
\begin{aligned}
L_1 ' \left( V_{i1} \right) &= \frac{- \mathrm{e}^{ -V_{i1} } }{1 +\mathrm{e}^{ -V_{i1} } }
\\
L_2 ' \left( V_{i1} \right) &=  \frac{1 }{1 + \mathrm{e}^{ - \gamma_1 V_{i1}}}
\\
P \left( y_{i1} = 1 \mid V_{i1} \right) &=  \frac{ L_{2}' \left( V_{i1} \right) }{ L_{2}' \left( V_{i1} \right) - L_{1}' \left( V_{i1} \right) }
\\
&= \frac{1}{1 - \frac{ L_{1}' \left( V_{i1} \right) }{L_{2}' \left( V_{i1} \right) }}
\\
&= \frac{1}{1 + \left( \dfrac{ 1 + \mathrm{e}^{ - \gamma_1 V_{i1}} }{ 1 +\mathrm{e}^{ -V_{i1} } } \right)  \mathrm{e}^{ -V_{i1} } }
\end{aligned}
\end{equation}
Because we derived this choice probability function from the uneven log-loss, we named it the ``uneven logit model.'' To visualize the range of possible shapes that the binary, uneven logit model can take, see Figure \ref{fig:asym_models}.

Now, using the procedure from Table \ref{table:procedure_for_creating_multinomial_extensions}, we can convert the choice probability function derived in Equation \ref{eq:binary_uneven_logit} into a logit-type model as given in Equation \ref{eq:logit_type_models}. We will omit the algebra, but the result is that we find $S \left( V_{ij}, \gamma_j \right) = V_{ij} + \ln \left( 1 + \mathrm{e}^{- V_{ij}} \right) - \ln \left( 1 + \mathrm{e}^{- \gamma _j V_{ij}} \right)$ and $\tau_j = 0 \ \forall j$. Note $\gamma_j$, for all alternatives $j$, is still required to be positive because this ensures that $S_{ij}$ is monotonically increasing in $V_{ij}$.

As with the multinomial clog-log and multinomial scobit models, we can immediately generalize the uneven logit model to a conditional uneven logit model. This is done simply by allowing $x_{i2} \neq 0$ and $\tau_j \neq 0$, although one of the $\tau _j$'s must still be constrained for identification purposes. Lastly, the multinomial uneven logit model is immediately obtained by using Equation \ref{eq:logit_type_models} with $S \left( \cdot \right)$ as derived in the last paragraph. As with the multinomial clog-log and scobit models, the choice probability function for the multinomial uneven logit model is displayed below for convenience.

\begin{equation*}
P_{\textrm{uneven logit}} \left( y_{ij} = 1 \mid \tau, \gamma, V_{i1}, V_{i2}, ..., V_{ik} \ \forall \left\lbrace j, k \right\rbrace \in C_i \right) = \frac{\exp \left[ \tau _j + V_{ij} + \ln \left( 1 + \mathrm{e}^{- V_{ij}} \right) - \ln \left( 1 + \mathrm{e}^{- \gamma _j V_{ij}} \right) \right]}{ \sum _{\ell \in C_i} \exp \left[ \tau _{\ell} + V_{i \ell} + \ln \left( 1 + \mathrm{e}^{- V_{i \ell}} \right) - \ln \left( 1 + \mathrm{e}^{- \gamma _j V_{i \ell}} \right) \right]}
\end{equation*}

\subsubsection{Example 4: Creating the Asymmetric Logit Model}
\label{sec:deriving_asym_logit}
Similar to Scott (\citeyear{scott_calibrated_2012}), Winkler (\citeyear{winkler_evaluating_1994}) in his paper ``Evaluating Probabilities: Asymmetric Scoring Rules'' developed a methodology for creating asymmetric loss functions from symmetric loss functions\footnote{Technically, Winkler developed a method for constructing asymmetric scoring rules from symmetric scoring rules. However, scoring rules are simply negated loss functions, so Winkler's methods also allow one to create asymmetric loss functions.}. However, unlike Scott, Winkler wanted to account for differing states of knowledge as opposed to different misclassification costs. In particular, Winkler wanted a loss function whose risk\footnote{Note the risk of a loss function is the expectation of the loss over all possible datasets, given the true parameters being estimated \citep{keener_statistical_2010}.} was maximized at the probability that corresponds to ``knowing nothing'' \citep{winkler_evaluating_1994}. This would allow one to judge probability forecasts in a way that accounts for the fact that knowing ``nothing'' does not always mean assigning a 50\% probability to the outcome of interest. Sometimes an analyst may still know that (on average) individuals have a greater or lesser than 50\% chance of choosing a given alternative. One such case where this is true is in class imbalanced situations. Given the link between Winkler's motivation for developing his asymmetric scoring rules and the class imbalanced scenarios that motivated this paper, we decided to investigate whether the choice probability functions derived from Winkler's asymmetric losses would be useful for making probability predictions under class imbalance. 

Applying Winkler's methods to the negative log-likelihood, and making the assumption that $V_{i2} = 0$, leads to the following asymmetric, negative log-likelihood:
\begin{equation}
\label{eq:asym_log_loss}
\begin{aligned}
\textrm{Asymmetric, Negative Log-Likelihood} &= 1_{\left\lbrace y_{i1} = 1 \right\rbrace} L_1 \left( P \left( y_{i1} = 1 \mid V_{i1}, V_{i2} = 0 \right) \right) +\\
&\quad \  1_{\left\lbrace y_{i1} = 0 \right\rbrace} L_2 \left( P \left( y_{i1} = 1 \mid V_{i1}, V_{i2} = 0 \right) \right)
\\
L_1 \left( P \left( y_{i1} = 1 \mid V_{i1}, V_{i2} = 0 \right) \right) &= \left\lbrace \begin{array}{cc}
\frac{\ln \left( \gamma_1 \right) - \ln \left[ P \left( y_{i1} = 1 \mid V_{i1}, V_{i2} = 0 \right) \right]}{- \ln \left( \gamma_1 \right)}, & P \left( y_{i1} = 1 \mid V_{i1}, V_{i2} = 0 \right) \geq \gamma_1 
\\[1.4ex]
\frac{\ln \left( \gamma_1 \right) - \ln \left[ P \left( y_{i1} = 1 \mid V_{i1}, V_{i2} = 0 \right) \right]}{- \ln \left( 1 - \gamma_1 \right)}, &\  P \left( y_{i1} = 1 \mid V_{i1}, V_{i2} = 0 \right) < \gamma_1
\end{array} \right\rbrace
\\[1.4ex]
L_2 \left( P \left( y_{i1} = 1 \mid V_{i1}, V_{i2} = 0 \right) \right) &= \left\lbrace \begin{array}{cc}
\frac{\ln \left( 1 - \gamma_1 \right) - \ln \left[ 1 - P \left( y_{i1} = 1 \mid V_{i1}, V_{i2} = 0 \right) \right]}{- \ln \left( \gamma_1 \right)}, & P \left( y_{i1} = 1 \mid V_{i1}, V_{i2} = 0 \right) \geq \gamma_1 
\\[1.4ex]
\frac{\ln \left( 1 - \gamma_1 \right) - \ln \left[ 1 - P \left( y_{i1} = 1 \mid V_{i1}, V_{i2} = 0 \right) \right]}{- \ln \left( 1 - \gamma_1 \right)}, & P \left( y_{i1} = 1 \mid V_{i1}, V_{i2} = 0 \right) < \gamma_1
\end{array} \right\rbrace \\
\textrm{where } \gamma_1 &\in \left( 0, 1 \right)
\end{aligned}
\end{equation}

Because the asymmetric, negative log-likelihood is piecewise defined, deriving the related choice probability function requires us to solve Equation \ref{eq:cpe_loss_to_binary_prob} twice, once for each case: $P \left( y_{i1} = 1 \mid V_{i1}, V_{i2} = 0 \right)$ greater than or equal to $\gamma_1$, and $P \left( y_{i1} = 1 \mid V_{i1}, V_{i2} = 0 \right)$ less than $\gamma_1$. The result will be a piecewise defined choice probability function. However, we need to avoid circular reasoning when defining the pieces of the choice probability function. In particular, we cannot define the pieces of the choice probability function using conditions based on the value of the choice probability function, as is done in the asymmetric, negative log-likelihood. To construct conditions for the related choice probability function, we note that Equation \ref{eq:cpe_loss_to_binary_prob} is a differential equation, so we will need boundary conditions to identify the constant of integration. Our boundary condition for the two cases will be that $P \left( y_{i1} = 1 \mid V_{i1} = 0, V_{i2} = 0 \right)$ equals $\gamma_1$. This condition will ensure continuity of the resulting choice probability function. Moreover, when combined with the fact that the choice probability function is monotonically increasing in $V_{i1}$, this boundary condition allows us to express the pieces of the choice probability function in terms of $V_{i1} \geq 0$ (which implies $P \left( y_{i1} = 1 \mid V_{i1}, V_{i2} = 0 \right) \geq \gamma_1$) and $V_{i1} < 0$ (which implies $P \left( y_{i1} = 1 \mid V_{i1}, V_{i2} = 0 \right) < \gamma_1$).

Starting with the case, $P \left( y_{i1} = 1 \mid V_{i1}, V_{i2} = 0 \right) \geq \gamma_1$ and $P \left( y_{i1} = 1 \mid V_{i1} =0, V_{i2} = 0 \right) = \gamma_1$, we have:
\begin{equation}
\begin{aligned}
\frac{d \left[ L_{2} \left( \hat{p} \left( V_{i1} \right) \right) \right]}{d \hat{p} \left( V_{i1} \right)} &= \frac{\hat{p} \left( V_{i1} \right) }{ \hat{p}' \left( V_{i1} \right) } \quad \textrm{where } \hat{p} \left( V_{i1} \right) = P \left( y_{i1} = 1 \mid V_{i1}, V_{i2} = 0 \right) \\
\left[ \frac{-1}{\ln \left( \gamma_1 \right)} \right] \frac{1}{1 - \hat{p} \left( V_{i1} \right)} &= \frac{\hat{p} \left( V_{i1} \right) }{ \hat{p}' \left( V_{i1} \right) } \\
\left[ \frac{-1}{\ln \left( \gamma_1 \right)} \right] \int \frac{ d \hat{p} }{\hat{p} \left( V_{i1} \right) \left[ 1 - \hat{p} \left( V_{i1} \right) \right]} &= \int d v \\
\left[ \frac{-1}{\ln \left( \gamma_1 \right)} \right] \ln \left( \frac{ \hat{p} \left( V_{i1} \right) }{ 1 - \hat{p} \left( V_{i1} \right) } \right) &= v + A \quad \textrm{where A is a constant} \\
\left[ \frac{-1}{\ln \left( \gamma_1 \right)} \right] \ln \left( \frac{ \hat{p} \left( V_{i1} \right) }{ 1 - \hat{p} \left( V_{i1} \right) } \right) &= v - \left[ \frac{1}{\ln \left( \gamma_1 \right)} \right] \ln \left( \frac{ \gamma_1 }{ 1 - \gamma_1 } \right)\\
\textrm{which simplifies to } \ \  \hat{p} \left( V_{i1} \right) &= \frac{1}{1 + \left( \gamma_1^{-1} - 1 \right) \gamma_1^{V_{i1}} }, \quad V_{i1} \geq 0,\  \gamma_1 \in \left( 0, 1 \right)
\end{aligned}
\end{equation}

Similarly, when $P \left( y_{i1} = 1 \mid V_{i1}, V_{i2} = 0 \right) < \gamma_1$ and $P \left( y_{i1} = 1 \mid V_{i1} =0, V_{i2} = 0 \right) = \gamma_1$, we have:
\begin{equation}
\begin{aligned}
\frac{d \left[ L_{2} \left( \hat{p} \left( V_{i1} \right) \right) \right]}{d \hat{p} \left( V_{i1} \right)} &= \frac{\hat{p} \left( V_{i1} \right) }{ \hat{p}' \left( V_{i1} \right) } \quad \textrm{where } \hat{p} \left( V_{i1} \right) = P \left( y_{i1} = 1 \mid V_{i1}, V_{i2} = 0 \right) \\
\left[ \frac{-1}{\ln \left( 1- \gamma_1 \right)} \right] \frac{1}{1 - \hat{p} \left( V_{i1} \right)} &= \frac{\hat{p} \left( V_{i1} \right) }{ \hat{p}' \left( V_{i1} \right) } \\
\left[ \frac{-1}{\ln \left( 1 - \gamma_1 \right)} \right] \int \frac{ d \hat{p} }{\hat{p} \left( V_{i1} \right) \left[ 1 - \hat{p} \left( V_{i1} \right) \right]} &= \int d v \\
\left[ \frac{-1}{\ln \left( 1 - \gamma_1 \right)} \right] \ln \left( \frac{ \hat{p} \left( V_{i1} \right) }{ 1 - \hat{p} \left( V_{i1} \right) } \right) &= v + B \quad \textrm{where B is a constant} \\
\left[ \frac{-1}{\ln \left( 1- \gamma_1 \right)} \right] \ln \left( \frac{ \hat{p} \left( V_{i1} \right) }{ 1 - \hat{p} \left( V_{i1} \right) } \right) &= v - \left[ \frac{1}{\ln \left( 1 - \gamma_1 \right)} \right] \ln \left( \frac{ \gamma_1 }{ 1 - \gamma_1 } \right)\\
\textrm{which simplifies to } \ \  \hat{p} \left( V_{i1} \right) &= \frac{1}{1 + \gamma _1 ^{-1} \left( 1 - \gamma_1 \right)^{ V_{i1} + 1} }, \quad V_{i1} < 0,\  \gamma_1 \in \left( 0, 1 \right)
\end{aligned}
\end{equation}

Together, the binary asymmetric logit model can be written as:
\begin{equation}
\label{eq:binary_asym_logit}
\begin{aligned}
P \left( y_{ij} = 1 \mid V_{i1}, V_{i2} = 0 \right) &= \left\lbrace \begin{array}{cc}
\dfrac{1}{1 + \left( \gamma_1^{-1} - 1 \right) \gamma_1^{V_{i1}} }, & V_{i1} \geq 0
\\[2ex]
\dfrac{1}{1 + \gamma _1 ^{-1} \left( 1 - \gamma_1 \right)^{ V_{i1} + 1} }, & V_{i1} < 0
\end{array} \right\rbrace
\end{aligned}
\end{equation}
For the readers' convenience, the binary asymmetric logit model is displayed in Figure \ref{fig:asym_models}, where we aim to highlight the range of possible shapes that the model can take. Note, we named this choice probability function the ``asymmetric logit model'' because it is derived from the asymmetric, negative log-likelihood. 

Now, following the procedure in Table \ref{table:procedure_for_creating_multinomial_extensions}, we can show that for the binary asymmetric logit model,
\begin{equation*}
\begin{aligned}
S \left( V_{ij}, \gamma _j \right) &= \left\lbrace \begin{array}{cc}
 \ln \left( \gamma _j \right) - V_{ij} \ln \left( \gamma_j \right) & V_{ij} \geq 0
 \\
 \ln \left( \gamma _j \right) - V_{ij} \ln \left( 1 - \gamma_j \right) & V_{ij} < 0
\end{array} \right\rbrace \\
\tau_j &= 0, \ \forall j
\end{aligned}
\end{equation*}
With these expressions, we can proceed from the binary case to the conditional and multinomial cases. In doing so, however, we must take care to generalize the restrictions and boundary condition used to derive the binary asymmetric logit model. In particular, we will require that
\begin{itemize}
\item $\gamma_j \in \left( 0, 1 \right) \ \forall j$,
\item $ \sum _j \gamma_j = 1$,
\item $ P \left( y_{ij} = 1 \mid V_{ik} = 0, \  \forall k \in C_i \right) = \gamma _j \ \forall j $, and 
\item that the multinomial, asymmetric logit model nest the multinomial logit model the same way the binary, asymmetric logit model nests the binary logit model\footnote{One can show that the binary, asymmetric logit model nests the standard binary logit model when $\gamma _j = \dfrac{1}{\parallel C_i \parallel}$.}.
\end{itemize}

With all of these requirements, the multinomial, asymmetric logit model can be written as given in Equation \ref{eq:logit_type_models}, where
\begin{equation}
\begin{aligned}
S \left( V_{ij}, \gamma _j \right) &= \left\lbrace \begin{array}{cc}
 \ln \left( \gamma _j \right) - V_{ij} \ln \left( \gamma_j \right), & V_{ij} \geq 0
 \\[2ex]
 \ln \left( \gamma _j \right) - V_{ij} \ln \left( \dfrac{1 - \gamma_j}{J - 1} \right), & V_{ij} < 0
\end{array} \right\rbrace \\
\textrm{where } J &= \textrm{The total number of possible alternatives in one's dataset}
\end{aligned}
\end{equation}
and where one of the $\tau_j$'s must be constrained for identification purposes. Note that unlike the multinomial clog-log, scobit, and uneven logit models, we will not display the choice probability function for the multinomial asymmetric logit model. Because of the piecewise definition of each $S_{ij}$, there are $2^J$ choice probability functions where $J$ is the total number of possible alternatives in the dataset. In other words, there is one function for each of the possible permutations of the indices $\left( V_{ij} \right)$ being positive or negative. Thus, even for three alternatives, we would need to display 8 equations. The simplest way of stating the multinomial asymmetric logit model is to refer to Equation \ref{eq:logit_type_models} and note that $S_{ij}$ is piecewise defined for all $j$ in this model.

\subsection{Summary}
\label{sec:logit_type_summary}
\begin{figure}
\centering
\includegraphics[width=\textwidth]{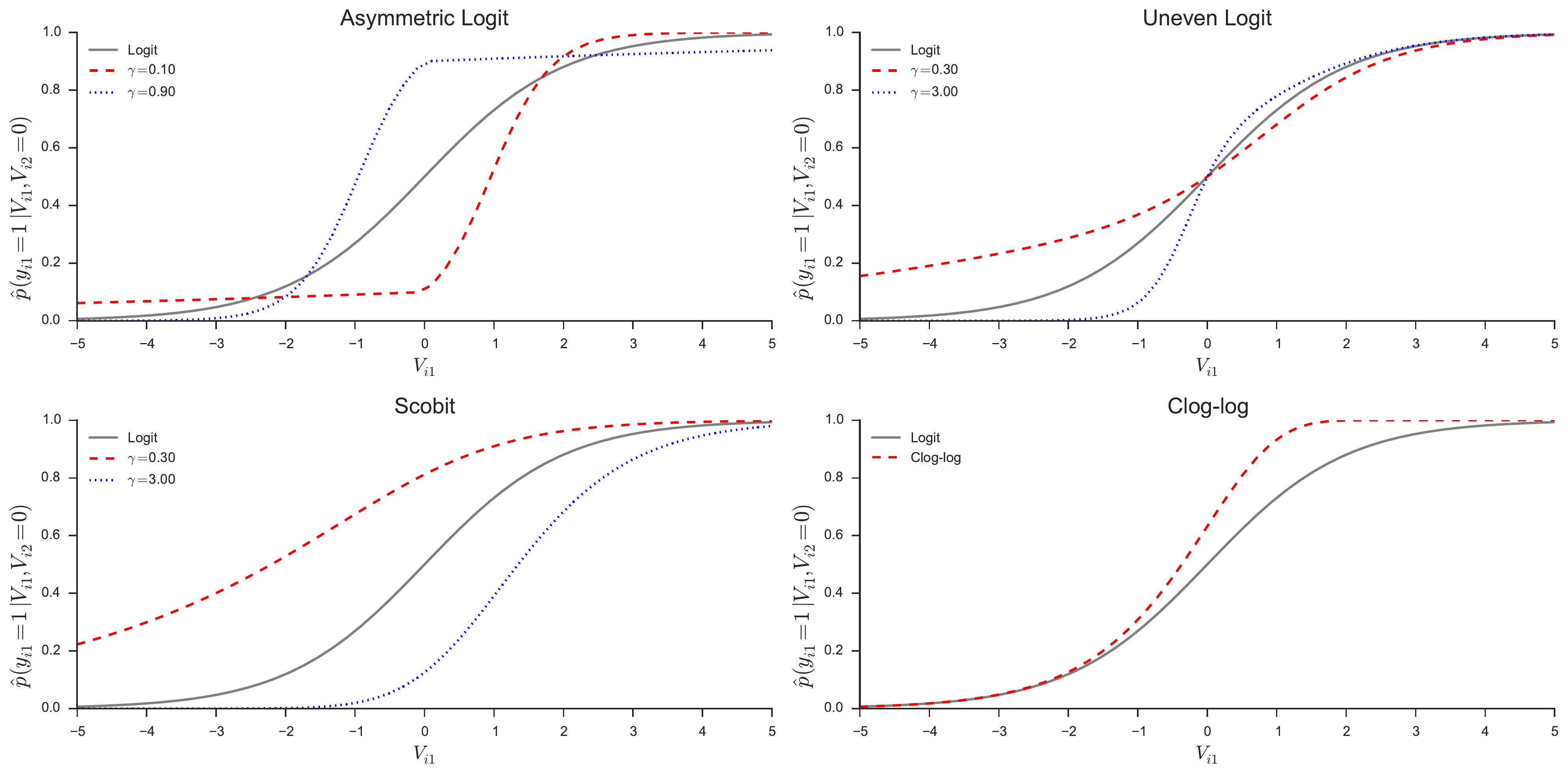}
\caption{Binary, Asymmetric Choice Models}
\label{fig:asym_models}
\end{figure}

To summarize, Section \ref{sec:logit_type_formulation} presented our proposed class of logit-type models and showed how they avoid the symmetry property described in the introduction. Section \ref{sec:logit_type_literature_relation} then positioned our logit-type models in relation to the existing discrete choice and statistics literature. Next, we showed in Section \ref{sec:binary_to_multinomial} how one can leverage the logit-type model formulation to extend existing, asymmetric choice models to the multinomial setting, thereby making such models useful to the transportation community at large. Finally, in Section \ref{sec:deriving_binary_asym_models}, we demonstrated one way to derive entirely new asymmetric choice models based on specific considerations that analysts may have concerning their study. Overall, we presented four new examples of this section's methods by deriving the multinomial clog-log, scobit, uneven logit, and asymmetric logit models. The binary versions of these models are shown in Figure \ref{fig:asym_models} to display the range of shapes that these models embody in comparison to the binary logit model. Note that we display the binary versions of these models instead of their multinomial versions simply for ease of visualization.

In the next Section, we will describe the estimation techniques used in this paper for the logit-type models given by Equation \ref{eq:logit_type_models}. Section \ref{sec:empirical_applications} will then present the empirical applications of our logit-type models and compare them to the standard multinomial logit model, using the four example models derived in this section. Section \ref{sec:extensions} will discuss extensions to our work and finally Section \ref{sec:conclusion} will conclude.

\section{Estimation Techniques}
\label{sec:model_estimation}
Within transportation, maximum likelihood estimation (MLE) is the most commonly used technique for performing statistical inference on the unknown parameters in one's discrete choice model. For the logit-type models specified in Equation \ref{eq:logit_type_models}, the gradient and hessian of the unknown parameters $\theta = \left( \beta, \tau, \gamma \right)$ can be calculated in closed-form, provided that $S\left( \cdot \right)$ is twice differentiable and provided that the unknown parameters are constrained such that $S\left( \cdot \right)$ exists. The existence of the gradient and hessian permits one to use most numeric optimization methods to try and maximize the likelihood of one's model. Even if $S\left( \cdot \right)$ is not differentiable, one may still be able to make use of sub-gradient methods to perform such numerical maximization. 

Despite having closed-form gradients and hessians, the log-likelihood of one's logit-type model will (in general) not be concave in the unknown parameters $\theta$. This lack of concavity can make it difficult to calculate the MLE estimates for one's logit-type model. Nevertheless, when possible, we used standard optimization techniques that do not require tuning parameters such as the Newton-Raphson or the Broyden-Fletcher-Goldfarb-Shanno (BFGS) algorithms. In cases where standard techniques failed, we resorted to custom-coded gradient descent algorithms. To implement the aforementioned estimation methods, all calculations were carried out using the Python programming language and the NumPy, SciPy, and Pandas packages \citep{mckinney_data_2010, van_numpy_2011, jones_scipy_2014}. Moreover, we developed a Python package called PyLogit to perform MLE for the MNL model and the four asymmetric models introduced in Section \ref{sec:logit_type_models}. Our package, PyLogit, is available for public use through the Python Package Index.

\section{Empirical Applications}
\label{sec:empirical_applications}

% Describe application
This section describes our two policy applications of the asymmetric choice models developed in this paper. These two applications were chosen because they differ in their respective emphases on the aggregate versus disaggregate predictions of the choice models. However, both applications use the same dataset and model specification. In particular, we model the travel mode choice of commuters in the San Francisco Bay Area who are making work or school tours. Our use of a common dataset and model specification allows us to consider the practical differences between the asymmetric and symmetric models based on use case, independent of differences in model inputs.

For our first application, we analyze the impact of a cordon toll in Downtown San Francisco on commute mode shares. As noted in the introduction, commute mode choices are almost always class imbalanced in the US. For instance, as shown in Table \ref{table:sample_mode_shares}, approximately 43\% of the 4,004 commute tours in our sample were conducted by driving alone while only 5\% were conducted by bicycling. In such class-imbalanced situations, it might be natural to suspect that one's choice probability function is asymmetric. We will investigate this hypothesis through statistical tests of our asymmetric choice models versus the MNL model and through each model's cross-validation performances. To evaluate the possible effects of the asymmetric choice models on policy analyses, in addition to the predictive performance of such models, we investigate the impact of cordon tolls on commute mode choice.

For our second application, we analyze the impact of using our asymmetric choice models in a travel demand management (TDM) setting. In particular, we focus on the use of individualized marketing to increase public transit ridership \citep{brog_individualized1998}. As a TDM strategy, individualized marketing targets individuals who do not currently use transit but nevertheless could be persuaded to use transit given the information and incentives being offered by the marketing campaign \citep{brog_individualized1998}. An example of one such incentive is the provision of free transit-passes for a limited time. This is the incentive used in our application. By assessing how ``switchable'' each individual is \citep{gensch_targeting_1984}, choice models such as the MNL model and the asymmetric models developed in this paper are used to select individuals for targeting and transit-pass provision. We then compare the costs and programmatic success of using the MNL model versus our asymmetric logit-type models for target selection in an individualized marketing campaign by treating our sample of individuals as the population of individuals that a transit agency's pilot marketing program might have access to. Together, the TDM and cordon toll analyses will provide insight into the nature of the practical differences between the asymmetric logit-type models and the traditional MNL model.

%In the subsection below, we will describe the dataset used in our two applications. Following this, we will describe the procedures we used for model estimation, model testing, our cordon toll analysis, and our TDM example. Finally, we will report our results, and conclude this section with a discussion.
In the subsection below, we report the main results of our model estimation and policy analyses. Following this, we conclude the section with a discussion. Readers who are interested in a detailed description of the data, surrounding context, and methodology used to conduct this analysis can see Appendix D (see Section \ref{appendix:data_and_methods}).

\begin{table}
\centering

\caption{Sample Mode Shares}
\label{table:sample_mode_shares}

\input{sample_mode_shares.tex}

\end{table}

\subsection{Results}
\label{sec:application_results}
In this sub-section, we report the results of our model estimation efforts for the standard MNL model and the four asymmetric choice models derived in this paper---the multinomial uneven logit, scobit, asymmetric logit, and clog-log models. The parameter estimates\footnote{As detailed in Appendix D, the parameter estimates for the shape parameters are reparameterized, and as such, are not the shape parameters described in Section \ref{sec:logit_type_models}. The reasons for the reparameterizations, as well as a precise description of them, are given in Appendix D.} are displayed in Table \ref{table:in_sample_MLE_estimation}. The asterisks that indicate statistical significance at the 95\% and 99\% confidence levels are based on the ``bias-corrected and accelerated'' (BCa) bootstrap confidence intervals of \citet{efron_introduction_1993} and \citet{diciccio_bootstrap_1996}. BCa intervals were used to assess statistical significance because our bootstrapping indicated that, at our current sample size, the sampling distributions of the MLEs for our asymmetric models had not yet converged to asymptotic normality. Since the sampling distributions of the MLEs had not converged to (approximate) asymptotic normality, the standard Wald tests based on such convergence were likely to be inaccurate \citep{jennings_judging_1986, pawitan_2000_reminder}. For full display of the 95\% and 99\% BCa intervals for each model, see Tables \ref{table:in_sample_MLE_interval_95} and \ref{table:in_sample_MLE_interval_99} in Appendix D.

Now, when interpreting the parameter estimation results displayed in Table \ref{table:in_sample_MLE_estimation}, one may wonder if the uneven logit model and asymmetric logit model are empirically identified. Indeed, for these two models, most of their shape parameters are not statistically significant at the 95\% confidence level. However, the uneven and asymmetric logit models are indeed empirically identified. As is well known in the statistics literature, if one is estimating both the shape parameters ($\gamma$) and the non-shape-parameters ($\tau$ and $\beta$) of a parametric link function (i.e. a choice probability function with parameters that control its shape), then the variance of one's estimates will be high when the shape parameters and non-shape-parameters are highly correlated \citep{stukel_generalized_1988, taylor_cost_1988, czado_orthogonalizing_1992}. As put by Czado, this variance increase is the ``cost'' of estimating the shape of the choice probability function within a particular parametric family \citep{czado_orthogonalizing_1992}. When we examined the correlation between shape ($\gamma$) and intercept parameters ($\tau$) of the uneven logit and asymmetric logit models, we found that these two sets of parameters were indeed highly correlated. This explains the non-significance of the estimated shape parameters and intercept terms. Moreover, each model's Hessian at the MLE had a small but existent curvature with respect to each model parameter, indicating that we do have empirical identification.

In addition to our parameter estimates, we also report the in-sample and out-of-sample predictive performance (i.e. the log-likelihoods) of each of the models in Tables \ref{table:in_sample_MLE_estimation}, \ref{table:mle_log_likelihood_by_mode}, and \ref{table:mle_cross_validation}. It can be seen that the asymmetric models generally had better predictive ability than the MNL model, both in-sample and out-of-sample. Finally, beyond the measures of statistical fit, we report the results of our applications on cordon pricing and individualized marketing for a public transportation TDM measure.

Specifically, for the cordon toll analysis, we report the aggregate, automobile-based mode share predictions by each choice model, in relation to the different toll amounts. These aggregate predictions are shown in Figure \ref{fig:mode_shares_by_toll}. Moreover, we present a comparison of the disaggregate probability forecasts of the MNL versus the uneven logit model in Figure \ref{fig:disagg_toll_prob_predictions} to highlight the disagreement between the asymmetric models and the MNL model. The map in Figure \ref{fig:drive_transit_walk_map} further emphasizes the practical significance of the differences between the MNL model and the asymmetric choice models used in this paper. 

Finally, for the individualized marketing application, our main results are shown in Figure \ref{fig:tdm_program_efficiencies}. Defining program efficiency as the total cost of providing the free transit-passes divided by the change in the expected number of walk-transit-walk riders, Figure \ref{fig:tdm_program_efficiencies} displays the program efficiencies that are achieved by using each of the choice models for target identification. As mentioned in the following discussion section, it is useful to have Table \ref{table:mle_log_likelihood_by_mode} to help assess the program efficiency results shown in Figure \ref{fig:tdm_program_efficiencies}. Table \ref{table:mle_log_likelihood_by_mode} decomposes the overall in-sample log-likelihoods achieved by each model into the in-sample log-likelihoods achieved on each travel mode. It allows us to compare the program efficiency results to the predictive ability of each model on specific travel modes instead of just interpreting the program efficiency results based on overall model performance. In general, all of the results mentioned above will be discussed more thoroughly in the discussion section to follow (Section \ref{sec:application_discussion}).

\subsection{Discussion}
\label{sec:application_discussion}

\subsubsection{Model Estimation and Testing}
As shown in Tables \ref{table:in_sample_MLE_estimation} and \ref{table:mle_cross_validation}, the multinomial clog-log model did not perform well relative to the MNL model. However, all of the asymmetric choice models with flexible shapes (i.e. with shape parameters) more accurately predicted the mode choice of individuals in our sample than the MNL model. In particular, there were large differences in in-sample log-likelihoods between the asymmetric choice models with flexible shapes and the MNL model. These differences range from about 132 for the asymmetric logit model to 205 for the uneven logit model. Since all three of the asymmetric choice models with shape parameters nest the MNL model, log-likelihood ratio tests were used to assess whether the differences in model fit were statistically significant. Table \ref{table:in_sample_MLE_estimation} shows that all three of the asymmetric, flexible shape models had log-likelihood ratio statistics that were significant at the 0.01 alpha-level. This suggests that the MNL model is inappropriate for this dataset, relative to the flexible, asymmetric choice models used in this paper. Moreover, the greater predictive ability of the uneven logit, the asymmetric logit, and the scobit model was not limited to just the in-sample predictions. The out-of-sample predictions in Table \ref{table:mle_cross_validation} showed exactly the same trends indicated by the in-sample results. Here, the differences in the average out-of-sample log-likelihood during cross-validation ranged from approximately 12 for the asymmetric logit to 20 for the uneven logit. Given that the testing sets in each fold of the cross-validation are about one-tenth the size of the overall dataset, these results are consistent with the in-sample results. This indicates that the greater predictive ability of the flexible, asymmetric choice models as compared to the MNL are real and not due to over-fitting.

% Report in-sample results
\begin{landscape}
\begin{table}
\centering

\caption{MLE Parameter Estimation Results}
\label{table:in_sample_MLE_estimation}

\input{in_sample_MLE_bootstrap_res.tex}

\end{table}
\end{landscape}

% Report the in-sample log-likelihoods by mode
\begin{table}
\centering

\caption{MLE In-Sample Log-likelihoods by Travel Mode and by Model}
\label{table:mle_log_likelihood_by_mode}

\input{in_sample_log_likelihoods_by_mode_Bootstrap.tex}

\end{table}

% Report out-of-sample results
\begin{table}
\centering

\caption{MLE Average Out-of-Sample Log-likelihood During 10-fold Cross-Validation}
\label{table:mle_cross_validation}

\input{mle_cross_validation_results.tex}

\end{table}

% Report policy analysis results
\begin{figure}
\centering
\includegraphics[width=\textwidth]{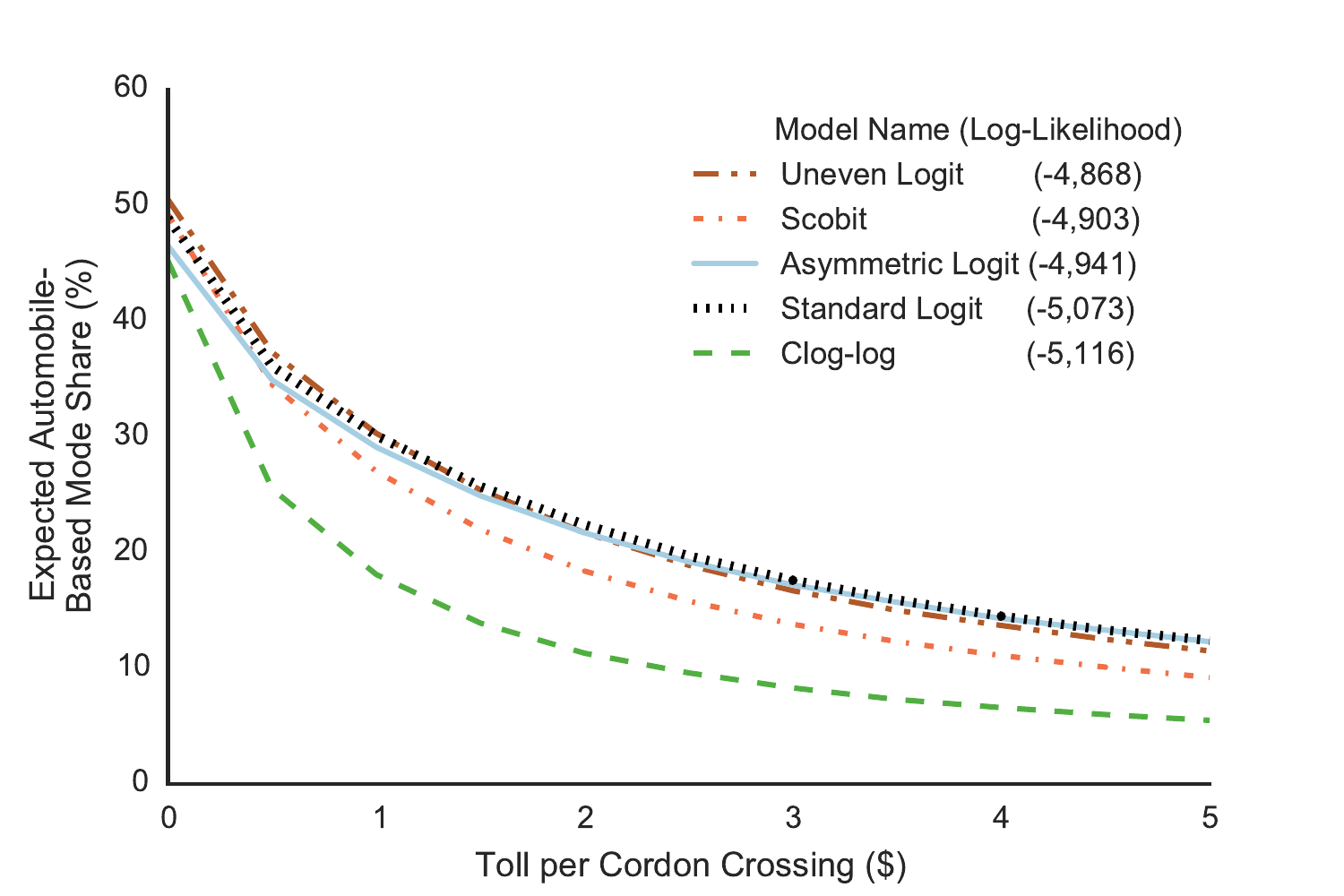}
\caption{Automobile-Based Mode Shares by Model and by Cordon Toll Amount}
\label{fig:mode_shares_by_toll}
\end{figure}

\begin{figure}
\centering
\includegraphics[width=\textwidth]{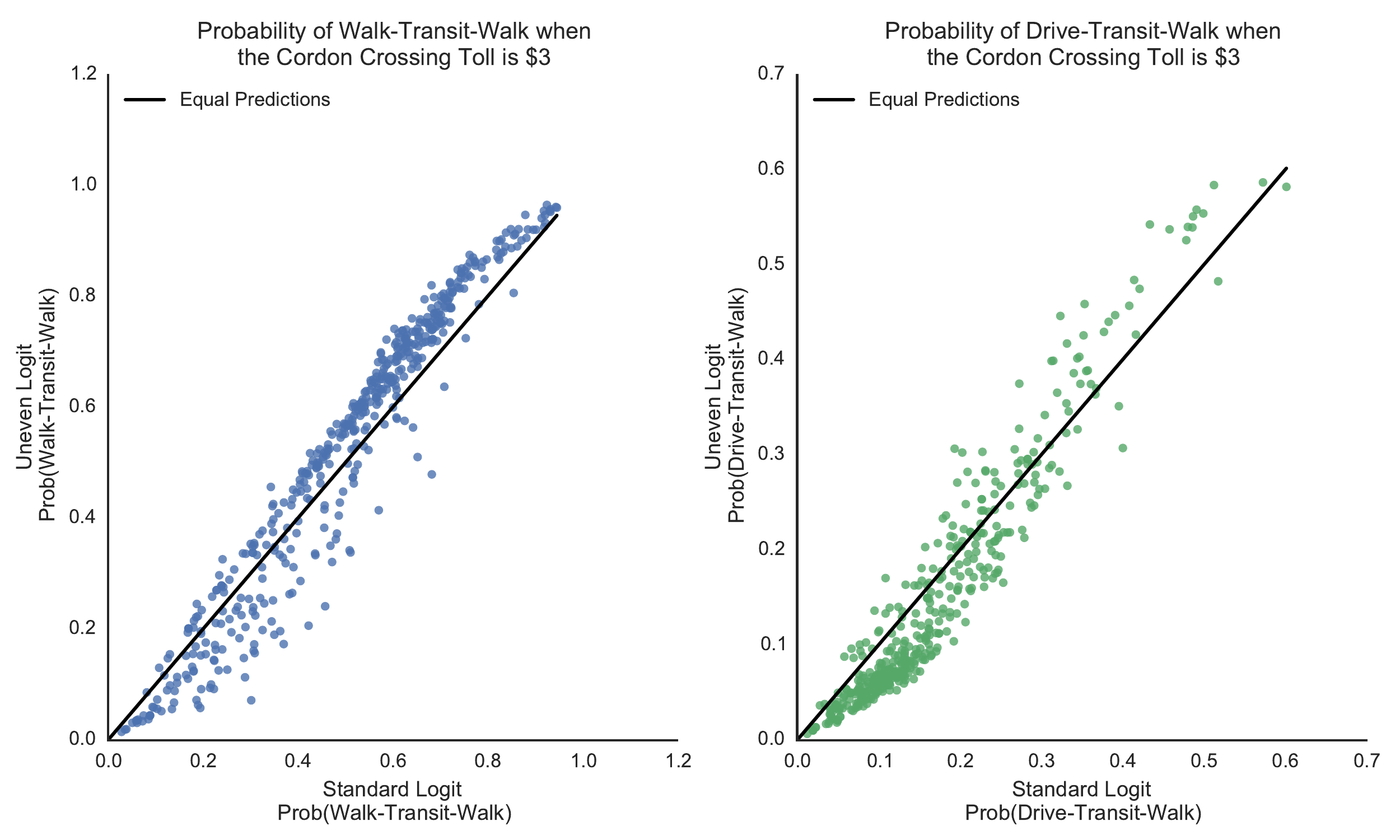}
\caption{Disaggregate Probability Predictions for Walk-Transit-Walk and Drive-Transit-Walk for the Uneven Logit and the MNL Models}
\label{fig:disagg_toll_prob_predictions}
\end{figure}

\begin{figure}
\centering
\includegraphics[width=\textwidth]{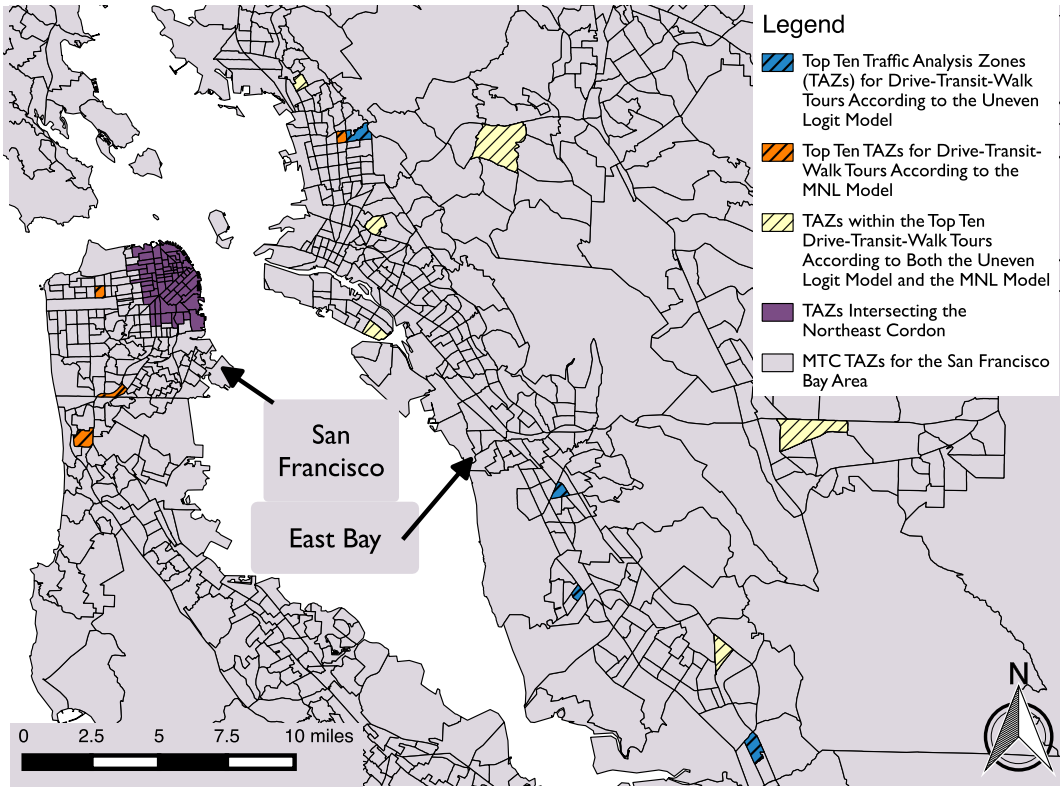}
\caption{Top Ten Traffic Analysis Zones Producing Drive-Transit-Tours \newline According to the Uneven Logit and MNL Model at \$3 per Cordon Crossing}
\label{fig:drive_transit_walk_map}
\end{figure}

\begin{figure}
\centering
\includegraphics[width=\textwidth]{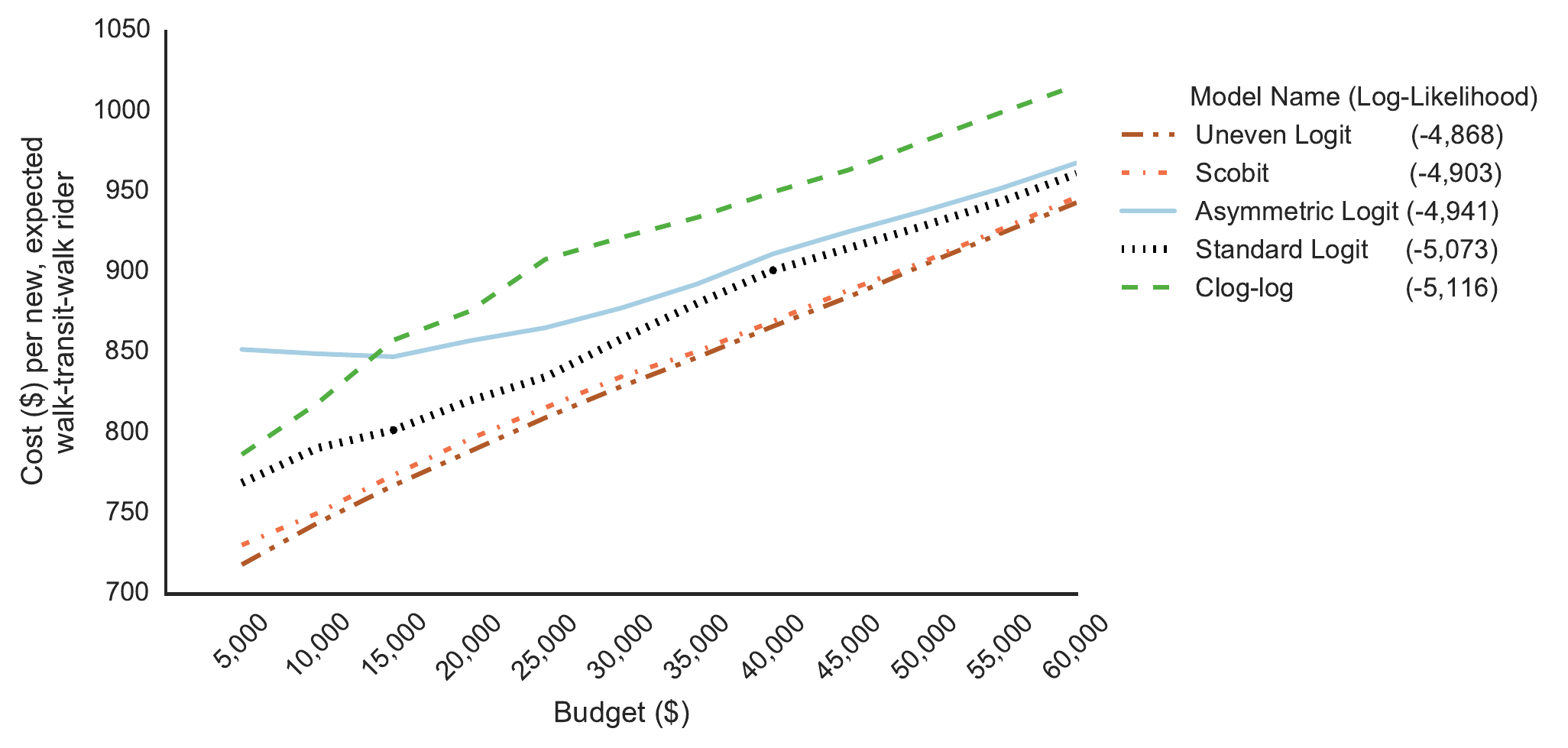}
\caption{Total Individualized Marketing Costs Per New, Expected Walk-Transit-Walk Rider by Model}
\label{fig:tdm_program_efficiencies}
\end{figure}

\subsubsection{Cordon Toll Analysis}
In addition to judging whether the improvements offered by one model over another are ``statistically significant," it is important to assess whether such improvements are ``practically significant.'' One way we assessed the practical impacts of the asymmetric choice models derived in this paper was to conduct an analysis of the effects of a congestion cordon toll\footnote{From economic theory, if a set of alternatives are perfect substitutes for one another, then the marginal dis-utility of cost should be constant across the alternatives since the goods are exactly the same. Because our estimated cost-coefficients for the Drive Alone, SharedRide-2, and SharedRide-3+ modes show large differences from one another, we have reason to believe that these three automobile alternatives are not perfect substitutes for one another, even after controlling for our study's explanatory variables. Accordingly, we conclude that there is some set of unobserved variables that still differentiates the three modes from each other and which interacts with the cost variable to influence an individual's cost-sensitivity for each mode. While recognizing this issue, we are not sure what these unobserved variables might be, and even if we did have thoughts about what these unobserved variables might be, we are not in a position to collect data on these features. As a result, both our cordon toll and TDM analysis are therefore conditional on the following \textit{ceteris paribus} assumption: that the interaction effects between cost and these unobserved variables remain as they currently are, despite external changes to the cost of the various automobile-based modes. Thanks to an anonymous reviewer for raising this point.} in Downtown San Francisco. 

At the most basic level, we compare the MNL model and the asymmetric choice models on the basis of their aggregate, predicted mode shares for automobile-based modes (drive alone, shared ride with two passengers, and shared ride with three or more passengers) under various cordon toll charges. Given that the purpose of the congestion toll is to reduce the use of automobile-based modes at peak commute times, large differences in predicted mode share for automobile-based modes would have great ramifications for support and expectations of the congestion tolling scheme. As shown in Figure \ref{fig:mode_shares_by_toll}, the aggregate mode share predictions for the automobile based modes, for tours that cross the cordon, follows the same general trend for both the MNL and the flexible, asymmetric choice models. Moreover, the differences in the predicted mode shares are minimal. Compared to the flexible, asymmetric models, the MNL model overestimates the mode share of automobile-based modes by 1-3.8\% at the San Francisco County Transportation Authority's proposed toll of \$3 per cordon crossing, depending on which model is being examined. However, the MNL and flexible asymmetric models all predicted overall decreases of approximately 31-35\% in automobile based mode shares from a \$0 toll to a \$3 toll. In light of the overall predicted mode share changes, the differences between models seems mostly inconsequential from a general planning perspective.

Beyond the basic question of how the aggregate, automobile-based mode shares will change as a result of tolling, a host of disaggregate outputs from mode choice models may be useful for transportation agencies implementing the congestion toll. In particular, to support individuals making their commute trips under the tolling scheme, transportation agencies should make switching to more sustainable modes (such as public transit) as easy and safe as possible. For example, at transit stations where one expects the average number of drive-transit-walk commuters to increase, and where parking capacity is nearly full at peak hours, transit agencies might want to increase parking capacity so that 'park-and-ride' trips can be more readily accommodated. However, such actions require knowledge of which transit stations have catchment areas that are going to see large increases in their drive-transit-walk mode shares. 

To be accurate, these station-level determinations require accurate predictions of the disaggregate drive-transit-walk probabilities for individuals. In our application, we find substantive disagreements between the MNL model and the flexible, asymmetric choice models. For example, Figure \ref{fig:disagg_toll_prob_predictions} shows the predicted probabilities of walk-transit-walk and drive-transit-walk with a cordon toll of \$3 according to both the MNL model and the uneven logit model for the 4,004 tours in our sample. As can be seen, many of these predicted probabilities disagree. These disagreements are not just an artifact of the \$3 toll, but they exist at every tolling amount we tested, including the base case scenario with no toll. The substantive impact of these individual-level disagreements is that practitioners deciding where to install pedestrian improvements and increase parking capacity based on the MNL model might make misguided decisions: installing infrastructure where it is not needed, or failing to install infrastructure where it is needed. For instance, Figure \ref{fig:drive_transit_walk_map} shows the ten traffic-analysis-zones producing the greatest expected numbers of drive-transit-walk tours into the cordon area for the MNL model and the Uneven Logit model at \$3 per cordon crossing. As shown by the map, the MNL model under-predicts the amount of drive-transit-walk trips from the East Bay into the cordon area, relative to the uneven logit model. Practitioners using the MNL model as opposed to the uneven logit model might then incorrectly underestimate the need for increased parking capacity at BART stations in the East Bay, thereby hampering the success of the congestion pricing effort.

\subsubsection{TDM Analysis---Individualized Marketing}
Continuing with the emphasis on disaggregate model differences, this subsection discusses the practical differences for an individualized marketing campaign for TDM. Here, we assume the role of an agency interested in maximizing the increase in the expected number of walk-transit-walk riders per dollars expended. As such, the differences that we are concerned with in this application result from selecting individuals for targeting using each of the choice models being compared in this paper. To the extent that the different models select different individuals, the costs of providing the transit-passes will differ, and the change in the expected number of new walk-transit-walk commuters will differ.

Using the uneven logit model to estimate the ``true'' change in the expected number of new walk-transit-walk commuters (since the uneven logit model had the highest in-sample and out-of-sample log-likelihoods---see Tables \ref{table:in_sample_MLE_estimation} and \ref{table:mle_cross_validation}), Figure \ref{fig:tdm_program_efficiencies} shows the ratio of the ``program efficiencies'' achieved by each model, for a range of budgets for purchasing the transit passes. From this Figure, a few insights can be gleaned. 

First, when only a small proportion of the sample can be targeted (i.e. when the budget is low), the scobit and uneven logit models make the best uses of money relative to the MNL model. For instance, with a \$5,000 budget, the MNL spends \$770 per new expected walk-transit-walk passenger, while the scobit and uneven logit models spend \$731 and \$719, respectively. If the number of individuals in the marketing program increases while the proportion that is targeted remains the same, such differences in program efficiency will lead to large differences in the number of new walk-transit-walk riders that are attracted using each model's targeting list. Second, as the budget increases and the proportion of individuals that can be targeted increases, the differences between the program efficiencies of each model are greatly reduced. This is to be expected. At the limit, there will be a large enough budget to select all individuals for targeting, thus the program costs and the ``true'' increase in the expected number of walk-transit-walk passengers will be equal across models. Lastly, the ranking of program efficiencies across models depended not on the overall predictive ability of one's multinomial model but mostly on the predictive ability of one's model for the travel mode of interest (walk-transit-walk in this case). For instance, since the asymmetric logit model's in-sample and out-of-sample log-likelihoods are higher than those of the MNL model (see Tables \ref{table:in_sample_MLE_estimation} and \ref{table:mle_cross_validation}), one would expect the asymmetric logit model to make better targeting selections than the MNL model. However, when one looks at the log-likelihoods of each model for just the walk-transit-walk mode (shown in Table \ref{table:mle_log_likelihood_by_mode}), we can see that the asymmetric logit model is actually a worse predictor of the walk-transit-walk mode than the MNL model, even though it is has a higher log-likelihood overall. It's program efficiency is therefore worse than that of the MNL model. Another seemingly anomalous fact is that when the budget is low, the clog-log model is able to better target individuals than the asymmetric logit model. This merely reflects the fact that for this sample and relative to the asymmetric logit model, the clog-log model is better able to find the small handful of individuals providing the highest increase in their probability of commuting via walk-transit-walk per dollar spent. However, as the budget increases and the number of individuals that is to be targeted increases, the ranking of program efficiencies return to the predictable state of mimicking the in-sample, log-likelihood rankings for the walk-transit-walk mode.

Overall, for our individualized marketing application, we find that when resources are limited (i.e. when only a small percentage of one's population can be targeted for marketing), the use of the MNL model can be inefficient as compared to the asymmetric choice models such as the uneven logit and scobit models. In our example, such inefficiencies cost the MNL an additional \$51 per new expected walk-transit-walk rider when compared to the uneven logit model. As the budget for the marketing campaign and the percentage of individuals that could be targeted increased, the disaggregate predictive abilities of each model became less important, and as with the cordon toll application, the practical differences between models became minimal.

\subsection{Summary}
Through our analysis of the commute mode choices of San Francisco Bay Area residents, we found that the three asymmetric choice models with flexible shapes (i.e. those with shape parameters) had much better predictive ability (overall) than the standard MNL model. This result was observed in both in-sample and out-of-sample log-likelihoods. Moreover, these results were corroborated through our log-likelihood ratio test results. All of our flexible asymmetric models had log-likelihoods that were higher than the MNL model, at statistically significant levels.

With regard to practically significant differences, we find that the MNL model and the flexible, asymmetric choice models yield similar \textit{aggregate} inferences in our cordon toll analysis. In particular, results concerning the aggregate mode shares of automobile-based modes at different tolling levels are virtually equal across the different models. The practical differences between the MNL and flexible, asymmetric models comes from the \textit{disaggregate} predictions of the various models for each individual, and the fact that the predictions for some modes may differ greatly. Specifically, the predictions for walk-transit-walk and drive-transit-walk differ greatly between the MNL model and the flexible, asymmetric models in our example. 

The practical significance of these differences for our cordon toll application is that discordant inferences are obtained regarding where transit-serving parking supply should be increased. The MNL model suggests that transit-serving parking should be added in San Francisco itself, whereas the asymmetric models imply that the most important places to increase transit-serving parking supply are all in the East Bay. Due to the higher land values in San Francisco and a lower supply of land to devote to parking, the use of the MNL model instead of its better performing, asymmetric counterparts would lead transportation agencies to misguidedly spend much more money providing parking in San Francisco, when the East Bay is likely in greater need for transit-serving parking under a congestion tolling scheme.

For our TDM application, the practical significance of our models is that the asymmetric choice models that predict the walk-transit-walk mode better than the MNL model can better guide investment of the money that is available for individualized marketing of public transit. Specifically, we found that in our example, when the budget for providing free transit passes was low (\$5,000), the cost of acquiring each new walk-transit-walk rider could be reduced by approximately \$50 and \$40, respectively, by using the uneven logit model and the scobit model for target selection instead of the MNL model. Conversely, as the budget and the percentage of individuals who could be targeted increased, the differences in the disaggregate predictions of the models mattered less and less for target selection and the resulting efficiency of the marketing campaign. This further underscores the fact that the practical usefulness of  asymmetric choice models appear to be highest when accurate, disaggregate predictions are needed.

Moving onto the remaining two sections of this paper, we will now transition from discussing our specific applications to looking more broadly at how our work on asymmetric, closed-form, finite-parameter models of multinomial choice can be extended. Then in Section \ref{sec:conclusion}, we will conclude by summarizing the theoretical contributions of this paper, highlighting our empirical results, and raising key research questions from this work that should interest academic scholars and professional analysts.

\section{Extensions}
\label{sec:extensions}
Thus far, all of the individual models and results that have been shown in this paper have been based on the general formulation of logit-type models given by Equation \ref{eq:logit_type_models}. Despite restricting ourselves to that proposed class of models, at least six extensions or future research directions are immediately apparent. In particular, these ideas for future work can be categorized as either (1) direct extensions of the logit-type models developed in this paper, (2) applications of this paper's ideas to other models, or (3) investigations of the statistical properties of logit-type models. In the following paragraphs we will detail each of the extensions and future research directions that comprise these categories.

Firstly, the logit-type models given in Equation \ref{eq:logit_type_models} can avoid the symmetry property of standard MNL models, but because they share the same functional form as the MNL model, they retain other undesired properties such as I.I.A. Accordingly, many of the motivations behind existing extensions to the MNL model remain equally applicable to our proposed class of logit-type models. Here we highlight three such extensions. First, models such as the ``Heteroskedastic Logit Model'' \citep{steckel_heterogeneous_1988, recker_discrete_1995, bhat_heteroscedastic_1995} allow the scale parameter to vary across observations, and this effectively allows the shape of the resulting choice probability function to vary across observations. An analogous extension to logit-type models would be to allow $\gamma_j$ to vary across individuals, such as by parametrizing it as a function of $x_{ij}$. Such parametrizations have been successfully used in a transportation context to improve the fit of binary scobit models \citep{zhang_scobit-based_2010, wu_analysis_2012}, but this type of extension can be more generally applied to any logit-type model that has shape parameters. Second, the wider class of ``multivariate extreme value''\footnote{This class was originally referred to as ``generalized extreme value'' (GEV) models \citep{mcfadden_econometric_1980}. The name multivariate extreme value was adopted to avoid confusion with the pre-existing generalized extreme value distribution \citep{jenkinson_frequency_1955}.} models (such as the nested and cross-nested logit) generalizes the MNL model, capturing arbitrary correlations between the utilities of an individual's various alternatives while still maintaining a closed-form expression \citep{train_discrete_2009}. Logit-type models would benefit from similar extensions. As mentioned in Section \ref{sec:logit_type_literature_relation}, one way to extend logit-type models to account for correlation between the utilities of one's alternatives is to specify various ``aggregation functions'' as described by Mattsson et al. (\citeyear{mattsson_extreme_2014}) in conjunction with $w_{ij} = \exp \left[ \tau_j + S \left( V_{ij}, \gamma_j \right) \right]$. Lastly, MNL models have been extended using various ``mixing distributions'' to account for taste heterogeneity in their parameters and to provide realistic substitution and correlation patterns between alternatives \citep{revelt_mixed_1998}. These mixed logit approaches use a MNL ``kernel'' and allow the $\beta$ coefficients to be randomly distributed throughout the population. Similar mixing strategies could be followed whereby one used a logit-type model as the kernel and a continuous mixing distribution of $\beta$s in the model. If using a discrete mixing distribution, i.e. a Latent Class Choice Model (LCCM), an analogous procedure is to use a logit-type model for the class-specific choice model. Such mixing procedures would allow for much greater flexibility and behavioral realism in our proposed logit-type models.

Beyond the direct extensions already mentioned, future research directions include applying the techniques and concerns of this paper to other choice models. Two such research directions seem immediately promising. First, as noted at the end of the last paragraph, one can consider using a logit-type model as the class-specific choice model in a LCCM. However, this still begs the question of what choice model should be used as the class-membership model. It is not clear that one would necessarily want the class-membership model to have the symmetry property described in the introduction, so it would be interesting to look at the effects of using an asymmetric, logit-type model as the class-membership model in one's LCCM. There could be large policy impacts from such a change. For instance, imagine one is interested in growing the market share of a desired market segment, such as a latent class of individuals with a predisposition towards using non-motorized modes of transportation. If that market segment is forecast to grow much more slowly when using an asymmetric model for the class membership probabilities as opposed to a MNL, and the asymmetric model fits one's data better, then policy-makers may need to take more aggressive measures to increase the market shares of the desired class. Secondly, the logit-type models developed in this paper were based on the desire to make the MNL model asymmetric. However, as stated above, this logit-based lineage leads to the inheritance of the other undesired properties of the logit model such as I.I.A. It would be interesting to instead try and make other, non-logit-based, choice models asymmetric. For instance, the Exponomial Choice model is not based on the logit model, yet it shares some of the attractive properties of the logit model. In particular, it has a closed-form probability equation, it has a log-likelihood that is concave with respect to the $\beta$s to be estimated, and it does not have the I.I.A. property \citep{alptekinoglu_exponomial_2016}. However, it is a symmetric choice probability function\footnote{This assertion is made based on plotting the choice probabilities for the binary exponomial choice model.}. It would be quite interesting to develop an asymmetric analogue to the Exponomial Choice model, as such a model would avoid both the I.I.A. property and the symmetry property.

Finally, there are a number of statistical questions regarding logit-type models that remain to be investigated. One point, raised by a referee, is that one does not know (a-priori) which logit-type model will be best for one's application. Therefore, one essentially has to try them all. As a result of this fact, it would be useful to study the characteristics of the best performing transformation functions $S \left( \cdot \right)$ in relation to the intrinsic characteristics of one's data. If such research leads to a greater understanding of how to specify one's $S \left( \cdot \right)$ functions, then one may be able to save researchers a fair amount of computational effort. More generally, it would be worthwhile to perform simulation studies to gain insight into the conditions under which asymmetric models perform better than symmetric ones and into the conditions that favor various types of asymmetry. In the meantime, the situation with logit-type models is analogous to the situation that is already faced in research, where (for instance) a researcher may not be certain (a-priori) of which plausible nesting or cross-nesting structure will be better in one's application.

Another statistical question is what is the best way to estimate one's logit-type model? As was noted in Section \ref{sec:model_estimation}, MLE was sometimes difficult for the four logit-type models derived in this paper. One response is to use Bayesian techniques to estimate the logit-type models since these techniques do not require maximization of an objective function. However, Bayesian estimation techniques can potentially lead to long estimation times, depending on one's model, dataset, and specific estimation method. It would be useful to investigate the properties of Bayesian and other estimation techniques on logit-type models. For instance, it has already been shown that maximum entropy estimation \citep{donoso_maximum_2011} may be a better estimation technique than MLE for nested logit models. Further research should be done with logit-type models to investigate the implications, the possible equivalences, and the relative merits and drawbacks of various estimation techniques such as bayesian inference, maximum entropy, method of moments, minimum chi-square estimation \citep{berkson_minimum_1980}, etc.

\section{Conclusion}
\label{sec:conclusion}

% Restate the main problem
% Restate one’s solution to the main problem
In this paper's introduction, we called attention to a symmetry property of common discrete choice models such as the MNL model and the simple probit model. Arguing that it is often undesirable for one's discrete choice model to a-priori be symmetric, we introduced a class of ``logit-type'' models that allow one to specify choice models of varying shapes and asymmetries, without entailing restrictions on the signs or magnitudes of the indices, $V_{ij}$.  Essentially, logit-type models replace the index, $V_{ij} = x_{ij} \beta$ in the MNL model with functions, $S \left( \cdot \right)$, that depend on the index and a finite number of shape parameters that control the shape of the choice probability function. By ensuring that this new function is asymmetric with respect to the index $V_{ij}$, we avoid symmetry in our logit-type models.

% Emphasize one’s contributions.
Next, we showed that our proposed class of models is both a parametrization of the class of models introduced by Mattsson et al. (\citeyear{mattsson_extreme_2014}) as well as a generalization of numerous existing, asymmetric choice models from both the transportation discipline as well as statistics. This nesting of existing models was used to devise a methodology for extending numerous pre-existing models to the ``conditional'' and multinomial settings. Such extensions greatly increase the number of situations that can be modeled by existing asymmetric choice models and increase the relevance of such models to transportation researchers whom often study inherently multinomial choice contexts. As examples of the proposed method, we extended two existing models---the clog-log model and the scobit model---to the multinomial setting for the first time.

Recognizing that the existing asymmetric choice models may not always suit a researcher's needs, we proposed a method for creating new, asymmetric choice models. We break from recent trends in transportation whereby one first specifies the distribution of each alternative's utility to each individual and then derives the choice probability functions as a result. Our paper takes the opposite approach of directly specifying the form of the choice probability functions, knowing that our logit-type models can be derived from innumerable distributions of the utilities. Doing so frees us to specify the choice probability functions according to the properties that we find desirable for our study. To demonstrate our proposed procedure, we derived two new choice models that generalize the MNL model: the asymmetric logit model and the uneven logit model.

To test the four new models derived in this paper against the standard MNL model, we applied all of these models to an analysis of travel mode choice in the San Francisco Bay Area. We find that all of the asymmetric choice models with flexible shapes (i.e. those with shape parameters to be estimated from the data) were able to fit the data better according to both in-sample and out-of-sample log-likelihoods. The difference in fit, for our example, was not just statistically significant but quite dramatic---on the order of more than 200 log-likelihood points for a dataset of only 4,004 individuals with 8 alternatives. Moreover, beyond the statistical fit and predictive ability of the various models, we showed that switching to asymmetric choice models can also entail serious policy implications. When looking at the effects of a cordon toll in Downtown San Francisco, we found that relative to the flexible asymmetric choice models (which had greater predictive power), the MNL model over-predicted the number of drive-transit-walk tours coming from San Francisco. Such over-predictions would encourage transportation agencies to erroneously invest more in increasing transit-serving parking supply in San Francisco as compared to the East Bay, where all of the other asymmetric models predict high expected numbers of drive-transit-walk tours. Moreover, in our TDM application, we find that the uneven logit model and the scobit model are able to better target individuals for marketing when funding for such a campaign is limited. In particular, the uneven logit and scobit models are able to reduce the cost of acquiring each new walk-transit-walk customer by approximately \$50 to \$40 relative to the MNL model when the marketing budget is only \$5,000. These results suggest that while asymmetric models may not always outperform symmetric ones, asymmetric choice models are at least worth testing in one's analysis as they might have better statistical performance and entail substantive policy and financial implications.

% Position one’s work in larger bodies of literature, both on the specific problem and in discrete choice and statistics at large
Lastly, while this paper presents a new class of models as well as four particular instances of this new class, many extensions to this work and future research directions remain. By direct analogy with MNL models, it will be of interest to extend logit-type models to account for arbitrary correlation structures between the various utilities of each alternative. Moreover, it will be interesting to make use of mixture formulations to incorporate taste heterogeneity and flexible patterns of substitution between alternatives. Regarding applications, further investigation remains to be done on the effect of incorporating logit-type models into other contexts (such as modeling market segmentation in LCCMs) and on the effect of incorporating asymmetry into choice models with different functional forms from the logit model (such as the Exponomial Choice model). Alongside all of the research directions mentioned above, there will of course be a need to answer statistical questions related to the proposed model-class, including questions of how best to estimate logit-type models and how one can check the appropriateness of a given function, $S \left( \cdot \right)$, for one's data.

\section*{Acknowledgements}
We thank James A. Goulet for many stimulating conversations in the beginning stages of this research. Additionally, we thank Michael Fratoni for computational assistance in the beginning of this project. Thanks go to Madeleine Sheehan and the anonymous referees for their constructive criticism of this manuscript. Any remaining errors or omissions are, of course, our own. Lastly, we thank UCCONNECT and Caltrans for funding this research effort.

\newpage
\section{Appendix A: Proofs}
\label{appendix:proofs}
Here we provide the derivation of Equation \ref{eq:cpe_loss_to_binary_prob}. It is based on Equation 16 and Equation 45 of \citet{buja_loss_2005}. Note that in all equations below, we use the notation introduced in Section \ref{sec:binary_to_multinomial}.

First, Equation 16 of \citeauthor{buja_loss_2005} states that:
\begin{equation}\tag{A1}
\label{eq:A1}
L_{2} \left( \hat{p} \left( V_{i1} \right) \right) = \int _0 ^{\hat{p}\left( V_{i1} \right)} tw\left(t\right)dt
\end{equation}
Applying the Fundamental Theorem of Calculus to Equation \ref{eq:A1}, we can write:
\begin{equation}\tag{A2}
\label{eq:A2}
\frac{d\left[ L_{2} \left( \hat{p} \left( V_{i1} \right) \right) \right]}{d\hat{p}} = \hat{p} \left( V_{i1}  \right) w \left( \hat{p} \left( V_{i1} \right) \right)
\end{equation}
At the same time, Equation 45 of \citeauthor{buja_loss_2005} states:
\begin{equation}\tag{A3}
\label{eq:A3}
1 = w \left( \hat{p} \left( V_{i1}  \right) \right) \hat{p}' \left( V_{i1}  \right)
\end{equation}
Assuming that $\hat{p}' \left( V_{i1}  \right) \neq 0$, we can rearrange Equation \ref{eq:A3} as follows:
\begin{equation}\tag{A4}
\label{eq:A4}
\frac{1}{\hat{p}' \left( V_{i1}  \right)} = w \left( \hat{p} \left( V_{i1}  \right) \right)
\end{equation}
Finally, substituting Equation \ref{eq:A4} into Equation \ref{eq:A2} yields Equation \ref{eq:cpe_loss_to_binary_prob}.

\newpage
\section{Appendix B: Further relations to existing literature}
\label{appendix:lit_relations}
In this appendix, we provide further details on the relationship between our models and those of \citet{li_multinomial_2011} and \citet{das_generalized_2014}. Additionally, for convenience, we provide a table showing how our proposed class of logit-type models subsumes previously described asymmetric models as special cases.

First, while the models based on Weibull, Rayleigh, Type II Generalized Logistic, Pareto, or Exponential distribution utilities are generalized by both our logit-type models and the single-index model of \citet{li_multinomial_2011}, our logit-type models given by Equation \ref{eq:logit_type_models} \textbf{\textit{are not}} a special case of Li's model. In particular, \citet{li_multinomial_2011} estimates a single $S \left( \cdot \right)$ for all alternatives. The transformations used in our logit-type models are allowed to differ across alternatives, based on the values of the shape parameters $\gamma _j$ for each alternative. Secondly, our paper provides a general method to create new, closed-form, binary probability functions. The paper of \citet{li_multinomial_2011} does not. \citet{li_multinomial_2011} instead focuses on semi-parametric, binary probability functions. Finally, our paper provides a method and rationale for generalizing binary probability functions to the multinomial setting, whether or not the distribution of the utilities underlying the probability function are known. The paper of \citet{li_multinomial_2011} provides no such distribution-free method for generalizing existing binary probability functions.

Next, with respect to the paper by \citet{das_generalized_2014}, our paper is strictly more general. \citet{das_generalized_2014} only consider a single $S\left( \cdot \right)$ function, namely that of \citet{czado_parametric_1994}. Our paper considers general, closed-form $S\left( \cdot \right)$ functions. As noted in Section \ref{sec:logit_type_literature_relation} and as shown below in Table \ref{table:logit-type-special-cases}, the model of \citet{das_generalized_2014} is a special case of the models described by our logit-type models given in Equation \ref{eq:logit_type_models}. Moreover, our paper provides a way to create such $S\left( \cdot \right)$ functions while the paper of \citet{das_generalized_2014} completely ignores this issue.

Lastly, we include Table \ref{table:logit-type-special-cases} in this appendix to explicitly show the various $S \left( \cdot \right)$ functions that show how a number of asymmetric probability functions from the literature can be seen as special cases of our logit-type models.

\begin{table}[h!]
\centering

\caption{Special Cases of Our Logit-Type Models}
\label{table:logit-type-special-cases}

\input{logit_type_special_cases.tex}

\end{table}

\newpage
\section{Appendix C: Deriving the MNL Model}
\label{appendix:deriving_mnl}

In this Appendix, we aim to clarify the procedures in Table \ref{table:procedure_for_creating_new_models} by deriving the familiar MNL model using both its related CPE loss and its related composite loss. We start with the composite loss, since it is a more straightforward derivation.

\paragraph{Deriving the Binary Logit Model Using the Log-Loss}
In Step 1, we are required to choose a binary loss function. For the binary logit model, one such loss function is the log-loss (i.e. the related composite loss for the binary logit model). As shown in Equation \ref{eq:log_loss}, the binary log-loss is:
\begin{equation*}
\begin{aligned}
\textrm{Log-Loss}\left(y_{i1},  V_{i1} \right) &= 1_{\left\lbrace y_{i1} = 1 \right\rbrace} \ln \left(1 + \mathrm{e}^{-V_{i1}} \right) + 1_{\left\lbrace y_{i1} = 0 \right\rbrace} \ln \left(1 + \mathrm{e}^{ V_{i1} } \right) \\
&= 1_{\left\lbrace y_{i1} = 1 \right\rbrace} L_1 \left( V_{i1} \right) + 1_{\left\lbrace y_{i1} = 0 \right\rbrace} L_2 \left( V_{i1} \right)
\end{aligned}
\end{equation*}

The necessary derivatives for Step 2 are $L_{1}' \left( V_{i1} \right)$ and $ L_{2}' \left( V_{i1} \right)$. For the log-loss, these derivatives are:
\begin{equation*}
\begin{aligned}
L_{1}' \left( V_{i1} \right) &= \frac{- \mathrm{e}^{-V_{i1}}}{1 + \mathrm{e}^{-V_{i1}}}\\
&= \frac{-1}{1 + \mathrm{e}^{V_{i1}}} \\
L_{2}' \left( V_{i1} \right) &=  \frac{ \mathrm{e}^{V_{i1}}}{1 + \mathrm{e}^{V_{i1}}}
\end{aligned}
\end{equation*}
Below, we use these derivatives to derive the formula for binary logit model that is commonly used in statistics and computer science applications (where $V_{i2}$ implicitly equals zero):
\begin{equation}\tag{C1}
\label{eq:binary_logit}
\begin{aligned}
P_{ \textrm{binary logit} } \left( y_{i1} = 1 \mid V_{i1} \right) &= \frac{ L_{2}' \left( V_{i1} \right) }{ L_{2}' \left( V_{i1} \right) - L_{1}' \left( V_{i1} \right) } \\
&= \dfrac{ \dfrac{ \mathrm{e}^{V_{i1}}}{1 + \mathrm{e}^{V_{i1}}}} { \dfrac{ \mathrm{e}^{V_{i1}}}{1 + \mathrm{e}^{V_{i1}}} - \dfrac{-1}{1 + \mathrm{e}^{V_{i1}}} } \\
&= \dfrac{ \dfrac{ \mathrm{e}^{V_{i1}}}{1 + \mathrm{e}^{V_{i1}}}} { \dfrac{ \mathrm{e}^{V_{i1}} + 1}{ 1 + \mathrm{e}^{V_{i1}}}  } \\
&= \dfrac{ \mathrm{e}^{V_{i1}} }{ 1 + \mathrm{e}^{V_{i1}} }
\end{aligned}
\end{equation}

For a moment, we will defer Step 3 (where we extend the binary logit model to the multinomial setting). Instead we will now show how the same  binary logit model formula can be obtained from the negative log-likelihood. The final step of extending the binary logit model to the MNL model will be the same for both versions of the procedure in Table \ref{table:procedure_for_creating_new_models}, regardless of whether we start with a CPE loss or a composite loss function.

\paragraph{Deriving the Binary Logit Model Using the Negative Log-Likelihood} Similar to the use of the log-loss, we can use the negative log-likelihood as our CPE loss function from which we derive the binary logit model. As given in Equation \ref{eq:neg_log_likelihood}, the negative log-likelihood with $V_{i2}$ assumed to equal zero is 
\begin{equation*}
\begin{aligned}
\textrm{Negative Log-Likelihood}\left(y_{i1},  P \left( y_{i1} = 1 \mid V_{i1}, V_{i2} = 0 \right) \right) &= 1_{\left\lbrace y_{i1} = 1 \right\rbrace} \left( - \ln \left[ P \left( y_{i1} = 1 \mid V_{i1}, V_{i2} = 0 \right) \right] \right) +\\
&\quad \  1_{\left\lbrace y_{i1} = 0 \right\rbrace} \left( - \ln \left[ 1 - P \left( y_{i1} = 1 \mid V_{i1}, V_{i2} = 0 \right) \right] \right) \\
&=  1_{\left\lbrace y_{i1} = 1 \right\rbrace} L_1 \left[ P \left( y_{i1} = 1 \mid V_{i1}, V_{i2} = 0 \right) \right] +\\
& \quad \ 1_{\left\lbrace y_{i1} = 0 \right\rbrace} L_2 \left[ P \left( y_{i1} = 1 \mid V_{i1}, V_{i2} = 0 \right) \right]
\end{aligned}
\end{equation*}

For Step 2, we need the derivative of $L_2$ with respect to $P \left( y_{i1} = 1 \mid V_{i1}, V_{i2} = 0 \right)$. This derivative is:
\begin{equation*}
\frac{\partial L_2 \left[ P \left( y_{i1} = 1 \mid V_{i1}, V_{i2} = 0 \right) \right]}{\partial P \left( y_{i1} = 1 \mid V_{i1}, V_{i2} = 0 \right)} = \frac{1}{1 - P \left( y_{i1} = 1 \mid V_{i1}, V_{i2} = 0 \right)}
\end{equation*}
From here, we can use Equation \ref{eq:cpe_loss_to_binary_prob} to solve for $P \left( y_{i1} = 1 \mid V_{i1}, V_{i2} = 0 \right)$ as follows. Let $\hat{p} \left( V_{i1} \right) = P \left( y_{i1} = 1 \mid V_{i1}, V_{i2} = 0 \right)$. Then,
\begin{equation*}
\begin{aligned}
\frac{\partial L_2 \left[ \hat{p} \left( V_{i1} \right) \right]}{\partial \hat{p} \left( V_{i1} \right)} &= \frac{\hat{p} \left( V_{i1} \right)}{\dfrac{\partial \hat{p} \left( V_{i1} \right)}{\partial V_{i1}}} \\
\frac{1}{1 - \hat{p} \left( V_{i1} \right)} &= \frac{\hat{p} \left( V_{i1} \right)}{\dfrac{\partial \hat{p} \left( V_{i1} \right)}{\partial V_{i1}}} \\
\frac{1}{\hat{p} \left( V_{i1} \right) \left[ 1 - \hat{p} \left( V_{i1} \right) \right]} &= \frac{\partial V_{i1}}{\partial \hat{p} \left( V_{i1} \right)} \\
\frac{\partial \hat{p} \left( V_{i1} \right)}{\hat{p} \left( V_{i1} \right) \left[ 1 - \hat{p} \left( V_{i1} \right) \right]} &= \partial V_{i1} \\
\left[ \frac{1}{\hat{p} \left( V_{i1} \right)} + \frac{1}{1 - \hat{p} \left( V_{i1} \right)} \right]\partial \hat{p} \left( V_{i1} \right) &= \partial V_{i1} \\
\int \left[ \frac{1}{\hat{p} \left( V_{i1} \right)} + \frac{1}{1 - \hat{p} \left( V_{i1} \right)} \right]\partial \hat{p} \left( V_{i1} \right) &= \int \partial V_{i1} \\
\ln \left[ \hat{p} \left( V_{i1} \right) \right] - \ln \left[ 1 - \hat{p} \left( V_{i1} \right) \right] &= V_{i1} + A \quad \textrm{where A is a constant of integration} \\
\ln \left[ \frac{\hat{p} \left( V_{i1} \right)}{1 - \hat{p} \left( V_{i1} \right)} \right] &= V_{i1} + A 
\end{aligned}
\end{equation*}
As with any differential equation, we need a boundary condition to be able to determine the value of $A$. A typical condition would be that $\hat{p} \left( V_{i1} = 0 \right) = 0.5$. With this boundary condition, $A = 0$ and we have
\begin{equation*}
\ln \left[ \frac{\hat{p} \left( V_{i1} \right)}{1 - \hat{p} \left( V_{i1} \right)} \right] = V_{i1}
\end{equation*}
Standard algebraic manipulation leads back to Equation \ref{eq:binary_logit} for the binary logit model where $V_{i2}$ is assumed to be zero.

\paragraph{Extending Binary Logit to MNL}
Finally, Step 3 of our procedure for creating new choice models is that we use the procedure from Section \ref{sec:binary_to_multinomial} to create a multinomial extension of the binary version of our model. This is done below. The labels to the right of the equations refer to the steps in Table \ref{table:procedure_for_creating_multinomial_extensions}.

\begin{equation}\tag{C2}
\label{eq:binary_logit_to_mnl_part_1}
\begin{aligned}
P_{\textrm{binary logit}}  \left( y_{ij} = 1 \mid x_{i1}, x_{i2} = 0, \beta \right) &= \dfrac{ \mathrm{e}^{V_{i1}} }{ 1 + \mathrm{e}^{V_{i1}} } \qquad &\textrm{\textit{Step 2a.}} \\ 
\dfrac{ \mathrm{e}^{V_{i1}} }{ 1 + \mathrm{e}^{V_{i1}} } &\equiv \frac{\exp \left( S_{i1} \right)}{\sum _{\ell = 1} ^2 \exp \left( S_{i \ell} \right)}  \qquad &\textrm{\textit{Step 2b.}}\\
\frac{1}{1 + \mathrm{e}^{- V_{i1}}} &= \frac{1}{1 + \mathrm{e}^{S_{i2} - S_{i1}}} \\
1 + \mathrm{e}^{- V_{i1}} &= 1 + \mathrm{e}^{S_{i2} - S_{i1}} \\
- V_{i1} &= S_{i2} - S_{i1} \\
- V_{i1} &= \tau_2 + S \left( V_{i2} \right) - \tau_1 - S \left( V_{i1} \right)
\end{aligned}
\end{equation}
Using the same arguments as in Section \ref{sec:binary_to_multinomial}, we find $S \left( V_{ij} \right) = V_{ij}$. Substituting this equality back into the last line of Equation \ref{eq:binary_logit_to_mnl_part_1}, we get:
\begin{equation}\tag{C3}
\begin{aligned}
- V_{i1} &= \tau_2 + V_{i2} - \tau_1 - V_{i1} \\
0 &= \tau_2 + V_{i2} - \tau_1 \\
0 &= \tau_2 - \tau_1 \quad \textrm{because $V_{i2} = 0$} \qquad &\textrm{\textit{Step 2c.}} \\
\tau_1 &= \tau_2 \qquad &\textrm{\textit{Step 2d.}}
\end{aligned}
\end{equation}
The two constants, $\tau_1$ and $\tau_2$ are not identified. Without loss of generality we can set $\tau_1$, and implicitly $\tau_2$, equal to zero. This establishes the binary logit model as a special case of our logit-type models. The generalization to the MNL model given in Equation \ref{eq:mnl_formula} follows by removing the restrictions on $\tau$ and using Equation \ref{eq:logit_type_models} with $S \left( V_{ij} \right) = V_{ij}, \forall j \in C_i$, subject to identification. Typically, researchers include an alternative specific constant in $x_{ij}$. Such a constant will cause a lack of identification with $\tau$ in the MNL model. In such conditions, one can set $\tau_j = 0 \  \forall j$, and the MNL formula from Equation \ref{eq:mnl_formula} is recovered exactly.

\newpage
\section{Appendix D: Application Data and Methodology}
\label{appendix:data_and_methods}
In this appendix, we describe both the dataset used in our two applications as well as the methodology used to conduct our analysis. Specifically, we describe the procedures used for model estimation, model testing, our cordon toll analysis, and our TDM example.

% Describe data
\subsection{Data}
\label{sec:application_data}
The data used in this example comes from the 2012 California Household Travel Survey (CHTS). The CHTS was a one day travel diary taken from a stratified sample of households throughout the state of California and portions of Nevada. The complete data collection effort is described in \citet{california_department_of_transportation_2010-2012_2013}. For this study, the overall sample was filtered to include just those individuals commuting to school or work in the San Francisco Bay Area.

Beyond filtering based on geography and trip-purpose, we post-processed the raw CHTS data to construct the final dataset used for model estimation. In particular, we combined the data on individual trips into tours, defined a ``chosen travel mode'' for each tour, determined the available travel modes for each tour, and assembled the level-of-service variables for each tour. For this study, we used the level-of-service (travel costs, times, and distance) estimates provided by the San Francisco Metropolitan Transportation Commission (MTC). As a result, the set of possible alternatives in our example was defined to be the same as the categories used by MTC. Specifically, eight travel mode alternatives were specified. There were three driving modes, each differentiated by the number of passengers: drive-alone, shared-ride with two passengers, and shared-ride with three or more passengers. There were also three transit modes, each differentiated by their access and egress modes: walk-transit-walk (where walking is used for access and egress), drive-transit-walk, and walk-transit-drive. Finally, there were two non-motorized modes: walking and bicycling. For each tour, the travel mode that was used for the longest distance was used as the ``chosen travel mode'' for that tour. 

After filtering and post-processing, the final dataset consisted of 4,004 home-based work or school tours made by 3,836 individuals (with no individual making more than two tours). The percentage of tours that had their chosen travel mode associated with each of the aforementioned alternatives is shown in Table \ref{table:sample_mode_shares}. As mentioned earlier, the proportion of tours associated with each alternative is highly unbalanced, ranging from a low of 1.3\% for the share of ``walk-transit-drive'' tours to a high of 42.8\% for drive-alone tours.

\subsection{Estimation and Testing Procedures}
\label{sec:application_procedures}
% Describe the transformations used in the MLE. 
In this subsection, we will describe the procedures we used to perform the estimation, testing, and application of the various logit-type models employed in our example. 

\subsubsection{Estimation}
\label{sec:application_estimation_procedures}
First, to actually perform the numerical optimization necessary for the MLE, the scobit, the uneven logit, and the asymmetric logit models were re-parametrized. In particular, the log-likelihood functions of the scobit and the uneven logit models were expressed in terms of $ \Upsilon _j = \ln \left( \gamma _j \right) \ \forall j$, and the log-likelihood function of the asymmetric logit model was expressed in terms of $\Phi _j$ where $\gamma _j = \dfrac{\exp \left( \Phi _j \right)}{\sum _k \exp \left( \Phi _k \right)} \ \forall j$. These re-parametrizations allowed for unconstrained optimization of $\Upsilon _j$ and $\Phi _j$, and it led to better estimation results when compared to performing constrained optimizations on the original $\gamma _j$'s. Accordingly, our shape parameter estimates for the scobit, uneven logit, and asymmetric logit models are presented in terms of $\Upsilon _j$ and $\Phi _j$, respectively.

\subsubsection{Inference}
\label{sec:application_inference_procedure}
As noted in Section \ref{sec:application_results}, we used the non-parametric bootstrap and `bias-corrected and accelerated' (BCa) intervals of \citet{efron_introduction_1993} and \citet{diciccio_bootstrap_1996} to assess the statistical significance of our estimated parameters. This was done because, at our current sample size the sampling distributions of the MLE for the asymmetric models had not yet converged to an approximately normal distribution. The 95\% and 99\% BCa intervals are shown in Table \ref{table:in_sample_MLE_interval_95} and Table \ref{table:in_sample_MLE_interval_99}, respectively.

\begin{landscape}
     \begin{table}
		\centering

		\caption{MLE 95\% Bias-Corrected and Accelerated Confidence Intervals}
		\label{table:in_sample_MLE_interval_95}

		\input{in_sample_MLE_interval_95.tex}
	\end{table}
\end{landscape}

\begin{landscape}
	\begin{table}
		\centering

		\caption{MLE 99\% Bias-Corrected and Accelerated Confidence Intervals}
		\label{table:in_sample_MLE_interval_99}

		\input{in_sample_MLE_interval_99.tex}
	\end{table}
\end{landscape}

\subsubsection{Testing}
\label{sec:application_testing_procedures}
In our application, we use two types of model-testing or comparison procedures. First we use ``in-sample'' testing and comparison where the same sample that is used to estimate our models is then used to compare one model against another. The second type of model comparison and testing procedures that we use is ``out-of-sample'' where we use one subset of observations to estimate our models and then test our models against a different subset of observations. Because the MNL model is a restricted version of the uneven logit, asymmetric logit, and scobit models, we use log-likelihood ratio tests as our in-sample tests to compare the MNL versus the uneven logit model, the MNL versus the asymmetric logit model, and the MNL versus the scobit model. For our out-of-sample comparisons, we compare all of the models against one-another using ten-fold, stratified cross-validation. For this technique, we separate our data into ten stratified random subsets\footnote{Stratification is used so that the proportions of tours associated with each travel mode are relatively constant across the subsets.}. Then we iterate through the ten subsets, one at a time, using the selected subset for testing and the other nine subsets for estimation. The models are then compared on the basis of their average log-likelihoods across the ten subsets used for testing.

\subsubsection{Cordon Toll Analysis}
\label{sec:application_cordon_analysis_procedures}
\begin{figure}
\centering
\includegraphics[width=0.4\textwidth]{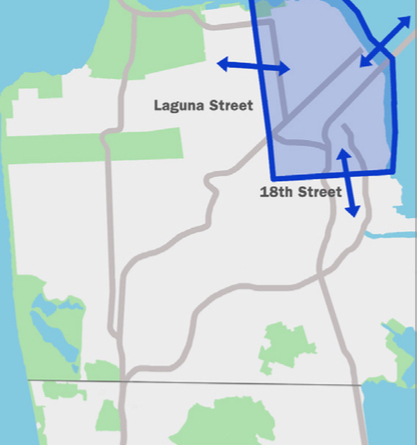}
\caption{Northeast Cordon for San Francisco Congestion Pricing \citep{sfcta_san_2010}}
\label{fig:northeast_cordon}
\end{figure}

% Describe the procedures used for the cordon toll analysis
The current congestion pricing proposal for the City of San Francisco is ``The Mobility, Access, and Pricing Study'' (MAPS) being conducted by the San Francisco County Transportation Authority (\citeyear{sfcta_san_2010}). The main congestion pricing alternative being studied is a \$3 toll that would be collected from cars passing into or out of the ``Northeast Cordon'' shown in Figure \ref{fig:northeast_cordon} during the AM peak (6AM - 10AM) or PM peak (3PM - 7PM), with individuals being charged no more than twice per day.

To study the effects of the proposed and similar congestion pricing schemes, we use sample enumeration based on the individual-level sample weights supplied by the CHTS. In particular, we varied the toll amount per crossing, from \$0 to \$5 in \$0.50 increments, calculated the probability of each travel mode for each tour given the current toll amount per crossing, and then used the sample weights to calculate the expected amount of tours using each mode. Care was taken to ensure that we properly calculated if, when, and how many times a tour would result in an individual driving into, out-of, or within the Northeast Cordon so that the toll could be applied as it is has been described in the MAPS study. 

Moreover, while it is unlikely that individuals will be using the walk-transit-drive or drive-transit-walk mode to commute into or out-of Downtown San Francisco (due to the lack of public parking lots at subway stations within the Northeast Cordon), we also applied the toll to those modes for people whose destination or origin (respectively) was within the cordon. Our rationale is that the purpose of the toll is to ease congestion within the Northeast Cordon. Using the walk-transit-drive or drive-transit-walk modes to commute into or out-of the Northeast Cordon is not supportive of such a purpose, even if one may not physically drive one's vehicle across the cordon. Our analysis therefore assumes that the agency implementing the congestion charge will devise a way to track and charge individuals using walk-transit-drive or drive-transit-walk to commute into or out-of locations inside the Northeast Cordon.

\subsubsection{TDM Analysis---Individualized Marketing}
\label{sec:application_tdm_analysis_procedures}
% Describe the procedures used for the TDM application
To better understand the differences between the standard MNL model and the asymmetric logit-type models developed in this paper, we asked the following question. Given a fixed budget to be spent on the provision of month-long free transit passes (such as for an individualized marketing pilot program), how would using the various asymmetric choice models for target selection compare to using the MNL model in terms of the dollars spent per expected new transit-rider?

To answer this question, we needed:
\begin{itemize}
\item a way to calculate the costs of transit-pass provision for each targeted individual,
\item a way to select individuals for targeting given the choice model being used, and
\item a way to assess each targeted individual's change in the probability of transit usage, given free transit.
\end{itemize}
First, we calculated the total cost of transit-pass provision for each individual by multiplying each individual's cost of the ``walk-transit-walk'' mode by an assumed 22 working days per month. Although one might typically take the cost of a month-long transit pass from relevant transit agencies and use this as the cost of transit-pass provision for each individual, transit agencies in the San Francisco Bay Area such as the Bay Area Rapid Transit (BART) System and Caltrain use distance-based fares. As a result, such agencies do not offer monthly passes, and we based our cost calculations on the individualized transit costs instead of using a single cost for all individuals. The idea is that each individual would be provided with a transit-pass that has been preloaded with the amount of money that is deemed necessary for the individual to complete one walk-transit-walk commute tour per working day for a month.

Secondly, to select individuals for targeting, we assumed the role of an agency that was interested in (1) incentivizing individuals to use the ``walk-transit-walk'' mode and (2) maximizing the increase in the expected number of walk-transit-walk riders per dollars expended. Based on these goals, our target selection procedure was as follows. We first calculate the probability of using the walk-transit-walk mode with and without a transit pass. Note that the provision of a transit pass would completely eliminate the cost of the walk-transit-walk mode, but it would also reduce the cost of the walk-transit-drive and drive-transit-walk modes by however much the individual would pay in walk-transit-walk costs\footnote{Of course, we assumed a minimum cost of \$0.}. Next we divide the change in the walk-transit-walk probability by the total cost of transit provision for each person. Finally, we place the individuals in descending order according to their change in walk-transit-walk probabilities per dollar spent, and we select all individuals from the top of the list such that the total cost of the transit-pass provision for all selected individuals is less than our specified budget. We repeated our analysis for a range of different budgets (\$5,000 - \$60,000) to better understand how the models perform in different scenarios.

Lastly, to assess each targeted individual's change in the probability of transit usage, we had to choose a model to treat as ``truth.'' As shown in Section \ref{sec:application_results}, the uneven logit model had the best in-sample and out-of-sample log-likelihoods. Given the dominant performance of the uneven logit model, we treated it as the ``true'' model that would be used to calculate the probability of an individual taking transit with or without a free transit-pass. Each model's probability predictions were therefore used to select the individuals for targeting as described in the last paragraph, but the uneven logit model was used when assessing the ratio of the total cost of the individualized marketing program to the total increase in the expected number of walk-transit-walk riders.

As one reviewer pointed out, treating the uneven logit model as the ``truth'' may be viewed as problematic because the shape parameters of the uneven logit model have confidence intervals that include zero (the same value as the standard MNL model). However, the joint confidence region of the vector of uneven logit shape parameters definitely excludes the vector of all zeros (i.e. the standard MNL model). This can be seen by the statistically significant results of the likelihood ratio test of the uneven logit model versus the standard MNL. As a result, even though the shape parameters have some uncertainty associated with them, the joint uncertainty is not so large as to include the standard MNL model. Because of these observations, and because we still need to be able to compare each model's predictions with some notion of `truth,' we treat the uneven logit model as the `truth.' Moreover, we choose to not include bootstrapped confidence intervals in Figure \ref{fig:tdm_program_efficiencies}. The point of this plot is to illustrate that differences between the various logit-type models are to be observed at the disaggregate level. This point would remain, even if overlapping confidence intervals were displayed, as the means of these confidence intervals would remain as they are now.

\newpage
\section*{\refname}
\bibliography{complete_draft_v8}

\end{document}

%% file: sample_mode_shares.tex
\begin{tabular}{lc}
\toprule
Travel Mode &  Mode Shares (\%) \\
\midrule
Drive Alone        &             42.8 \\
Shared Ride-2      &             15.9 \\
Shared Ride-3+     &             14.0 \\
Walk-Transit-Walk  &             10.3 \\
Drive-Transit-Walk &              \hphantom{*}1.5 \\
Walk-Transit-Drive &              \hphantom{*}1.3 \\
Walk               &              \hphantom{*}9.4 \\
Bike               &              \hphantom{*}4.6 \\
\bottomrule
\multicolumn{2}{l}{Note: Percentages do not sum to 100 due to rounding error.}
\end{tabular}

%% file: in_sample_MLE_bootstrap_res.tex
\begin{tabular}{K{0.275\linewidth}rrrrr}
\toprule
{}
Variables & \multicolumn{1}{c}{Standard Logit} & \multicolumn{1}{c}{Uneven Logit} & \multicolumn{1}{c}{Scobit} & \multicolumn{1}{c}{Asymmetric Logit} & \multicolumn{1}{c}{Clog-log} \tabularnewline
\midrule

\multicolumn{6}{l}{Alternative Specific Constants}\\
\quad Shared Ride: 2 & -1.010*\hphantom{*} & -0.806\hphantom{*}\hphantom{*} & -0.280\hphantom{*}\hphantom{*} & -1.242** & 0.969*\hphantom{*}\\
\quad Shared Ride: 3+ & 3.462** & 0.443\hphantom{*}\hphantom{*} & 2.596** & -0.724\hphantom{*}\hphantom{*} & 6.316*\hphantom{*}\\
\quad Walk-Transit-Walk & -0.392\hphantom{*}\hphantom{*} & 0.350\hphantom{*}\hphantom{*} & 11.524*\hphantom{*} & 0.490\hphantom{*}\hphantom{*} & -1.741**\\
\quad Drive-Transit-Walk & -2.622** & -3.002** & 4.388\hphantom{*}\hphantom{*} & 0.443\hphantom{*}\hphantom{*} & -4.001**\\
\quad Walk-Transit-Drive & -2.977** & -3.686** & 2.566\hphantom{*}\hphantom{*} & 0.451\hphantom{*}\hphantom{*} & -4.345**\\
\quad Walk & 1.554** & 1.626** & 0.156\hphantom{*}\hphantom{*} & 0.852** & -0.117\hphantom{*}\hphantom{*}\\
\quad Bike & -1.106** & -0.957** & -2.669** & 0.211*\hphantom{*} & -2.903**\\

\multicolumn{6}{l}{Travel Time, units:min}\\
\quad All Auto Modes & -0.076** & -4.376e-06** & -0.046** & -0.042** & -0.078**\\
\quad All Transit Modes & -0.027** & -0.364** & -0.003** & -0.016** & -0.026**\\

\multicolumn{6}{l}{Travel Cost}\\
\quad Units:\$ All Transit Modes & -0.127** & -1.718** & -0.015** & -0.080** & -0.210**\\
\quad Units:\$/mi Drive Alone & -5.061** & -3.718e-04** & -4.701** & -2.465*\hphantom{*} & -10.955**\\
\quad Units:\$/mi SharedRide-2 & -20.319** & -0.001** & -11.941** & -7.859** & -47.736*\hphantom{*}\\
\quad Units:\$/mi SharedRide-3+ & -90.922** & -0.002** & -32.494** & -16.531** & -141.947*\hphantom{*}\\

\multicolumn{6}{l}{Travel Distance, units:mi}\\
\quad Walk & -1.027** & -0.852** & -2.090** & -0.444** & -0.982**\\
\quad Bike & -0.287** & -0.211** & -0.465** & -0.164** & -0.263**\\

\multicolumn{6}{l}{Systematic Heterogeneity}\\
\quad Autos per licensed drivers (All Auto Modes) & 1.213** & 6.204e-05** & 0.597** & 0.452** & 0.764**\\
\quad Cross-Bay Tour (Shared Ride 2 \& 3+) & 0.928** & 7.841e-05** & 0.906** & 0.549** & 1.707**\\
\quad Household Size (Shared Ride 2 \& 3+) & 0.114*\hphantom{*} & 9.474e-06** & 0.074** & 0.053** & 0.073\hphantom{*}\hphantom{*}\\
\quad Number of Kids in Household (Shared Ride 2 \& 3+) & 0.687** & 3.587e-05** & 0.327** & 0.248** & 0.682**\\

\multicolumn{6}{l}{Shape Parameters}\\
\quad Drive Alone & \_\hphantom{*}\hphantom{*} & 9.716\hphantom{*}\hphantom{*} & 0.503** & \_\hphantom{*}\hphantom{*} & \_\hphantom{*}\hphantom{*}\\
\quad Shared Ride: 2 & \_\hphantom{*}\hphantom{*} & 10.000\hphantom{*}\hphantom{*} & 0.804** & 2.009** & \_\hphantom{*}\hphantom{*}\\
\quad Shared Ride: 3+ & \_\hphantom{*}\hphantom{*} & 10.190\hphantom{*}\hphantom{*} & 0.987** & 2.806\hphantom{*}\hphantom{*} & \_\hphantom{*}\hphantom{*}\\
\quad Walk-Transit-Walk & \_\hphantom{*}\hphantom{*} & -2.469\hphantom{*}\hphantom{*} & 2.917** & -1.342\hphantom{*}\hphantom{*} & \_\hphantom{*}\hphantom{*}\\
\quad Drive-Transit-Walk & \_\hphantom{*}\hphantom{*} & -2.820\hphantom{*}\hphantom{*} & 2.565*\hphantom{*} & -3.584\hphantom{*}\hphantom{*} & \_\hphantom{*}\hphantom{*}\\
\quad Walk-Transit-Drive & \_\hphantom{*}\hphantom{*} & -2.935\hphantom{*}\hphantom{*} & 2.434*\hphantom{*} & -3.953*\hphantom{*} & \_\hphantom{*}\hphantom{*}\\
\quad Walk & \_\hphantom{*}\hphantom{*} & 0.146\hphantom{*}\hphantom{*} & -0.811*\hphantom{*} & -0.958\hphantom{*}\hphantom{*} & \_\hphantom{*}\hphantom{*}\\
\quad Bicycle & \_\hphantom{*}\hphantom{*} & 0.279\hphantom{*}\hphantom{*} & -0.662\hphantom{*}\hphantom{*} & -1.632\hphantom{*}\hphantom{*} & \_\hphantom{*}\hphantom{*}\\

\multicolumn{6}{c}{}\tabularnewline

Log-Likelihood Ratio Stat. & \_\hphantom{*}\hphantom{*} & 410.148** & 341.273** & 264.828** & \_\hphantom{*}\hphantom{*}\\

Log-Likelihood & -5,073.428 & -4,868.354 & -4,902.791 & -4,941.014 & -5,116.066\\

\bottomrule
\multicolumn{6}{l}{Note: * means $\textrm{p-value} < 0.05$ and ** means $\textrm{p-value} < 0.01$.}
\end{tabular}

%% file: in_sample_log_likelihoods_by_mode_Bootstrap.tex
\begin{tabular}{lrrrrr}
\toprule
{} &  Standard Logit &  Uneven Logit &    Scobit &  Asymmetric Logit &  Clog-log \\
\midrule
Drive Alone        &      -1,084.14 &    -1,045.70 &  -1,040.07 &        -1,045.04 &  -1,092.69 \\
Shared Ride: 2     &      -1,183.66 &    -1,138.00 &  -1,144.92 &        -1,151.32 &  -1,196.55 \\
Shared Ride: 3+    &        -905.80 &      -826.15 &    -844.65 &          -847.57 &    -926.98 \\
Walk-Transit-Walk  &        -572.69 &      -566.87 &    -569.84 &          -572.85 &    -581.04 \\
Drive-Transit-Walk &        -184.99 &      -177.76 &    -177.20 &          -182.35 &    -185.73 \\
Walk-Transit-Drive &        -176.93 &      -167.30 &    -167.32 &          -176.54 &    -175.32 \\
Walk               &        -520.07 &      -502.27 &    -515.38 &          -519.54 &    -513.11 \\
Bike               &        -445.16 &      -444.30 &    -443.42 &          -445.80 &    -444.65 \\
\tabularnewline
Total              &       -5,073.43 &     -4,868.35 & -4,902.79 &         -4,941.01 & -5,116.07 \\
\bottomrule
\end{tabular}

%% file: mle_cross_validation_results.tex
\begin{tabular}{lc}
\toprule
Model &  Log-Likelihood \\
\midrule
Uneven Logit     &         -490.12 \\
Scobit           &         -494.04 \\
Asymmetric Logit &         -498.44 \\
Standard Logit   &         -510.28 \\
Clog-log         &         -514.63 \\
\bottomrule
\end{tabular}

%% file: logit_type_special_cases.tex
\begin{tabular}{K{0.325\linewidth}K{0.25\linewidth}cc}
\toprule
\multicolumn{1}{c}{Model} &  \multicolumn{1}{c}{$S \left( \cdot \right)$} & Shape Parameters & Constraints\\
\midrule
Exponential &  $- \log \left( V_{ij} \right)$ & N/A & $ V_{ij} > 0$\\
Rayleigh & $- 2 \log \left( V_{ij} \right)$ & N/A & $ V_{ij} > 0$ \\
Weibull &  $- \gamma \log \left( V_{ij} \right)$  & $\gamma$ & $ V_{ij}, \gamma > 0$\\
Pareto & $ \log \left( V_{ij} \right) - \log \left( V_{ij} - 1 \right)$ & N/A & $ V_{ij} > 1$ \\
Type II Generalized Logistic & $ \log \left[ \psi ^{-1}  \left( \psi \left( 1 \right) - \psi \left( V_{ij} \right) \right) \right]$ & N/A & $V_{ij} \notin \left\lbrace 0, -1, -2, ... \right\rbrace$ \\
\citet{das_generalized_2014} & $$\begin{array}{c}\frac{\left( 1 + V_{ij} \right)^{\gamma_{1j}} - 1}{\gamma_{1j}} \textrm{ if } V_{ij} \geq 0 \\[1.5ex] -\frac{\left( 1 - V_{ij} \right)^{\gamma_{2j}} - 1}{\gamma_{2j}} \textrm{ if } V_{ij} < 0 \end{array}$$ & $\gamma_j = \left[ \gamma_{1j}, \gamma_{2j} \right] \forall j$ & N/A\\
q-GEV \citep{nakayama_unified_2015} & $\frac{1}{1 - \gamma} \log \left[ 1 + \left( \gamma - 1 \right) V_{ij} \right]$ & $\gamma$ & $V_{ij} > \frac{-1}{\gamma - 1}$\\
\bottomrule
\multicolumn{4}{l}{Note $\psi \left( x \right) = \tfrac{\partial \log \left[ \Gamma \left( x \right) \right]}{\partial x }$ where $\Gamma \left( x \right)$ is the gamma function and N/A means ``not applicable.''}
\end{tabular}

%% file: in_sample_MLE_interval_95.tex
\begin{tabular}{K{0.275\linewidth}rrrrrrrrrr}
\toprule
{} & \multicolumn{2}{c}{Standard Logit} & \multicolumn{2}{c}{Uneven Logit} & \multicolumn{2}{c}{Scobit} & \multicolumn{2}{c}{Asymmetric Logit} & \multicolumn{2}{c}{Clog-log}\\
Variables & \multicolumn{1}{c}{2.5\%} & \multicolumn{1}{c}{97.5\%} & \multicolumn{1}{c}{2.5\%} & \multicolumn{1}{c}{97.5\%} & \multicolumn{1}{c}{2.5\%} & \multicolumn{1}{c}{97.5\%} & \multicolumn{1}{c}{2.5\%} & \multicolumn{1}{c}{97.5\%} & \multicolumn{1}{c}{2.5\%} & \multicolumn{1}{c}{97.5\%} \tabularnewline
\midrule

\multicolumn{11}{l}{Alternative Specific Constants}\\
\quad Shared Ride: 2 & -1.931 & -0.047 & -1.430 & 0.589 & -0.876 & 0.357 & -2.221 & -0.103 & 0.204 & 1.668\\
\quad Shared Ride: 3+ & 1.576 & 5.520 & -0.248 & 3.199 & 1.813 & 3.577 & -1.659 & 1.667 & 3.787 & 8.868\\
\quad Walk-Transit-Walk & -0.978 & 0.162 & -0.599 & 1.097 & 2.767 & 152.961 & -0.786 & 1.087 & -2.176 & -1.313\\
\quad Drive-Transit-Walk & -3.182 & -2.099 & -4.038 & -2.157 & -1.651 & 98.255 & -0.852 & 0.646 & -4.416 & -3.598\\
\quad Walk-Transit-Drive & -3.530 & -2.458 & -4.726 & -2.733 & -2.964 & 86.590 & -0.320 & 0.696 & -4.759 & -3.935\\
\quad Walk & 0.921 & 2.167 & 0.401 & 2.225 & -0.559 & 0.887 & 0.382 & 1.393 & -0.619 & 0.347\\
\quad Bike & -1.749 & -0.506 & -2.330 & -0.185 & -3.720 & -1.521 & 0.059 & 0.664 & -3.448 & -2.431\\

\multicolumn{11}{l}{Travel Time, units:min}\\
\quad All Auto Modes & -0.088 & -0.064 & -0.219 & -1.433e-07 & -0.058 & -0.035 & -0.048 & -0.035 & -0.090 & -0.065\\
\quad All Transit Modes & -0.032 & -0.022 & -0.543 & -1.372e-09 & -0.007 & -2.283e-04 & -0.019 & -0.014 & -0.031 & -0.022\\

\multicolumn{11}{l}{Travel Cost}\\
\quad Units:\$ All Transit Modes & -0.198 & -0.057 & -3.356 & -4.771e-08 & -0.045 & -0.001 & -0.113 & -0.037 & -0.284 & -0.131\\
\quad Units:\$/mi Drive Alone & -7.788 & -2.430 & -30.563 & -1.745e-05 & -7.662 & -2.792 & -3.918 & -0.281 & -13.308 & -8.666\\
\quad Units:\$/mi SharedRide-2 & -29.275 & -10.988 & -89.650 & -3.643e-05 & -16.506 & -8.174 & -11.680 & -3.460 & -54.964 & -39.743\\
\quad Units:\$/mi SharedRide-3+ & -119.584 & -64.001 & -164.792 & -7.217e-05 & -39.785 & -24.566 & -24.861 & -6.904 & -177.254 & -108.096\\

\multicolumn{11}{l}{Travel Distance, units:mi}\\
\quad Walk & -1.139 & -0.906 & -1.321 & -1.098e-04 & -4.743 & -1.058 & -0.566 & -0.212 & -1.093 & -0.865\\
\quad Bike & -0.345 & -0.224 & -10.468 & -3.629e-07 & -0.825 & -0.268 & -0.197 & -0.107 & -0.317 & -0.206\\

\multicolumn{11}{l}{Systematic Heterogeneity}\\
\quad Autos per licensed drivers (All Auto Modes) & 0.918 & 1.512 & 1.872e-06 & 1.982 & 0.444 & 0.783 & 0.346 & 0.612 & 0.527 & 0.991\\
\quad Cross-Bay Tour (Shared Ride 2 \& 3+) & 0.347 & 1.521 & 2.801e-06 & 5.112 & 0.596 & 1.260 & 0.257 & 0.827 & 1.056 & 2.350\\
\quad Household Size (Shared Ride 2 \& 3+) & 0.019 & 0.207 & 3.761e-07 & 0.301 & 0.030 & 0.121 & 0.018 & 0.092 & -0.021 & 0.169\\
\quad Number of Kids in Household (Shared Ride 2 \& 3+) & 0.573 & 0.806 & 1.076e-06 & 1.269 & 0.258 & 0.414 & 0.199 & 0.297 & 0.556 & 0.791\\

\multicolumn{11}{l}{Shape Parameters}\\
\quad Drive Alone & \_ & \_ & -1.153 & 13.071 & 0.191 & 0.707 & \_ & \_ & \_ & \_\\
\quad Shared Ride: 2 & \_ & \_ & -0.917 & 13.360 & 0.521 & 1.022 & 0.306 & 2.571 & \_ & \_\\
\quad Shared Ride: 3+ & \_ & \_ & -0.602 & 13.593 & 0.767 & 1.209 & -4.252 & 3.251 & \_ & \_\\
\quad Walk-Transit-Walk & \_ & \_ & -2.844 & 17.100 & 1.820 & 5.406 & -4.908 & 0.601 & \_ & \_\\
\quad Drive-Transit-Walk & \_ & \_ & -3.240 & 16.587 & 1.528 & 5.018 & -4.315 & 3.418 & \_ & \_\\
\quad Walk-Transit-Drive & \_ & \_ & -3.396 & 16.319 & 1.388 & 4.898 & -4.806 & -1.146 & \_ & \_\\
\quad Walk & \_ & \_ & -2.464 & 6.929 & -1.688 & -0.046 & -3.461 & 0.515 & \_ & \_\\
\quad Bicycle & \_ & \_ & -3.502 & 13.289 & -1.517 & 0.166 & -2.885 & 1.539 & \_ & \_\\

\bottomrule
\end{tabular}

%% file: in_sample_MLE_interval_99.tex
\begin{tabular}{K{0.275\linewidth}rrrrrrrrrr}
\toprule
{} & \multicolumn{2}{c}{Standard Logit} & \multicolumn{2}{c}{Uneven Logit} & \multicolumn{2}{c}{Scobit} & \multicolumn{2}{c}{Asymmetric Logit} & \multicolumn{2}{c}{Clog-log}\\
Variables & \multicolumn{1}{c}{0.5\%} & \multicolumn{1}{c}{99.5\%} & \multicolumn{1}{c}{0.5\%} & \multicolumn{1}{c}{99.5\%} & \multicolumn{1}{c}{0.5\%} & \multicolumn{1}{c}{99.5\%} & \multicolumn{1}{c}{0.5\%} & \multicolumn{1}{c}{99.5\%} & \multicolumn{1}{c}{0.5\%} & \multicolumn{1}{c}{99.5\%} \tabularnewline
\midrule

\multicolumn{11}{l}{Alternative Specific Constants}\\
\quad Shared Ride: 2 & -2.248 & 0.280 & -1.649 & 0.922 & -1.088 & 0.599 & -2.939 & -0.103 & -3.343 & 1.883\\
\quad Shared Ride: 3+ & 0.870 & 6.222 & -0.495 & 4.144 & 1.556 & 3.867 & -2.589 & 1.717 & -4.710 & 9.747\\
\quad Walk-Transit-Walk & -1.151 & 0.341 & -1.003 & 1.210 & -0.781 & 169.288 & -2.052 & 1.087 & -2.321 & -1.118\\
\quad Drive-Transit-Walk & -3.348 & -1.912 & -4.557 & -2.025 & -4.432 & 102.822 & -3.067 & 0.715 & -4.540 & -3.409\\
\quad Walk-Transit-Drive & -3.693 & -2.279 & -5.255 & -2.569 & -5.391 & 97.877 & -1.226 & 0.696 & -4.878 & -3.747\\
\quad Walk & 0.682 & 2.358 & 0.057 & 2.397 & -0.766 & 1.200 & 0.289 & 1.393 & -0.773 & 0.547\\
\quad Bike & -2.016 & -0.298 & -2.627 & -0.009 & -4.059 & -0.446 & -0.870 & 0.813 & -3.595 & -2.241\\

\multicolumn{11}{l}{Travel Time, units:min}\\
\quad All Auto Modes & -0.091 & -0.061 & -0.325 & -4.046e-08 & -0.063 & -0.030 & -0.049 & -0.030 & -0.093 & -0.047\\
\quad All Transit Modes & -0.034 & -0.021 & -0.656 & -1.404e-10 & -0.022 & -2.271e-04 & -0.020 & -0.013 & -0.033 & -0.020\\

\multicolumn{11}{l}{Travel Cost}\\
\quad Units:\$ All Transit Modes & -0.218 & -0.031 & -4.610 & -2.79e-09 & -0.131 & -0.001 & -0.124 & -0.021 & -0.307 & -0.032\\
\quad Units:\$/mi Drive Alone & -8.777 & -1.677 & -75.446 & -3.344e-06 & -8.770 & -2.207 & -4.445 & 0.325 & -14.158 & -4.659\\
\quad Units:\$/mi SharedRide-2 & -32.749 & -7.675 & -134.983 & -9.149e-06 & -18.268 & -7.097 & -13.312 & -2.554 & -57.399 & 6.917\\
\quad Units:\$/mi SharedRide-3+ & -129.730 & -56.617 & -212.323 & -1.994e-05 & -42.701 & -21.672 & -28.281 & -4.896 & -189.148 & 15.216\\

\multicolumn{11}{l}{Travel Distance, units:mi}\\
\quad Walk & -1.182 & -0.877 & -11.685 & -4.183e-09 & -4.781 & -0.822 & -0.586 & -0.141 & -1.133 & -0.834\\
\quad Bike & -0.365 & -0.206 & -12.898 & -6.228e-09 & -1.385 & -0.166 & -0.210 & -0.060 & -0.334 & -0.191\\

\multicolumn{11}{l}{Systematic Heterogeneity}\\
\quad Autos per licensed drivers (All Auto Modes) & 0.798 & 1.600 & 5.799e-07 & 3.144 & 0.399 & 0.861 & 0.302 & 0.676 & 0.277 & 1.068\\
\quad Cross-Bay Tour (Shared Ride 2 \& 3+) & 0.197 & 1.671 & 6.597e-07 & 7.344 & 0.491 & 1.385 & 0.161 & 0.926 & 0.400 & 2.515\\
\quad Household Size (Shared Ride 2 \& 3+) & -0.015 & 0.235 & 6.513e-08 & 0.633 & 0.014 & 0.138 & 0.006 & 0.102 & -0.053 & 0.197\\
\quad Number of Kids in Household (Shared Ride 2 \& 3+) & 0.538 & 0.847 & 3.086e-07 & 1.829 & 0.238 & 0.449 & 0.184 & 0.318 & 0.171 & 0.832\\

\multicolumn{11}{l}{Shape Parameters}\\
\quad Drive Alone & \_ & \_ & -1.553 & 14.357 & 0.061 & 0.766 & \_ & \_ & \_ & \_\\
\quad Shared Ride: 2 & \_ & \_ & -1.335 & 14.653 & 0.407 & 1.090 & 0.306 & 4.346 & \_ & \_\\
\quad Shared Ride: 3+ & \_ & \_ & -0.968 & 14.870 & 0.696 & 1.292 & -4.252 & 4.447 & \_ & \_\\
\quad Walk-Transit-Walk & \_ & \_ & -3.025 & 19.054 & 0.348 & 5.506 & -4.908 & 1.802 & \_ & \_\\
\quad Drive-Transit-Walk & \_ & \_ & -3.456 & 18.545 & -0.020 & 5.047 & -4.736 & 71.050 & \_ & \_\\
\quad Walk-Transit-Drive & \_ & \_ & -3.625 & 18.176 & -0.190 & 5.005 & -5.115 & 18.627 & \_ & \_\\
\quad Walk & \_ & \_ & -2.806 & 8.506 & -1.754 & 0.268 & -5.751 & 0.786 & \_ & \_\\
\quad Bicycle & \_ & \_ & -3.739 & 16.642 & -2.399 & 0.692 & -4.870 & 66.272 & \_ & \_\\

\bottomrule
\end{tabular}